\documentclass[11pt]{article}

\usepackage[margin=1in]{geometry}
\usepackage{amsmath,mathtools,amsthm,amssymb}
\usepackage{parskip}
\usepackage{graphicx}
\usepackage{courier}
\usepackage{caption}
\usepackage{booktabs}

\usepackage{multirow}
\usepackage{enumitem}
\usepackage{array}
\usepackage[ruled]{algorithm2e}
\usepackage{lipsum}
\usepackage{xcolor}
\usepackage{hyperref}
\usepackage{color}
\usepackage{rotating}
\usepackage{pdflscape}

\usepackage{graphics, epsfig, amsfonts, latexsym, adjustbox, ltablex, fancyhdr}
\fancypagestyle{plain}{
 \fancyhf{}
 \fancyfoot[C]{\iffloatpage{}{\thepage}}
 }
\newtheorem{theorem}{Theorem}

\newtheorem{proposition}{Proposition}

\newtheorem{assumption}{Assumption}

\usepackage{natbib}
\usepackage{array}
\usepackage{soul}

\newcommand\independent{\protect\mathpalette{\protect\independenT}{\perp}}
\def\independenT#1#2{\mathrel{\rlap{$#1#2$}\mkern2mu{#1#2}}}

\title{Partial identification and unmeasured confounding with multiple treatments and multiple outcomes}

\author{Suyeon Kang\footnote{Assistant Professor, School of Data, Mathematical, and Statistical Sciences, University of Central Florida (Email: suyeon.kang@ucf.edu)},
~~Alexander Franks\footnote{Associate professor, Department of Statistics and Applied Probability, University of California, Santa Barbara (Email: afranks@pstat.ucsb.edu)},
~~Heejun Shin\footnote{Postdoctoral Research Fellow, Department of Biostatistics, Harvard T.H. Chan School of Public Health (Email: heejunshin@hsph.harvard.edu)},
\\
Michelle Audirac\footnote{Senior Data Science Specialist, Department of Biostatistics, Harvard T.H. Chan School of Public Health (Email: maudirac@hsph.harvard.edu)},
~~Danielle Braun\footnote{Principal research scientist, Department of Biostatistics, Harvard T.H. Chan School of Public Health, Department of Data Science, Dana-Farber Cancer Institute (Email: dbraun@mail.harvard.edu)},
~~Joseph Antonelli\footnote{Associate Professor, Department of Statistics, University of Florida (Email: jantonelli@ufl.edu)}}

\date{}

\begin{document}

\maketitle

\begin{abstract}
Estimating the health effects of multiple air pollutants is a crucial problem in public health, but one that is difficult due to unmeasured confounding bias. Motivated by this issue, we develop a framework for partial identification of causal effects in the presence of unmeasured confounding in settings with multiple treatments and multiple outcomes. Under a factor confounding assumption, we show that joint partial identification regions for multiple estimands can be more informative than considering partial identification for individual estimands one at a time. We show how assumptions related to the strength of confounding or magnitude of plausible effect sizes for one estimand can reduce the partial identification regions for other estimands. As a special case of this result, we explore how negative control assumptions reduce partial identification regions and discuss conditions under which point identification can be obtained. We develop novel computational approaches to finding partial identification regions under a variety of these assumptions. We then estimate the causal effect of PM$_{2.5}$ components on a variety of public health outcomes in the United States Medicare cohort, where we find that the detrimental effect of certain air pollutants are robust to the potential presence of unmeasured confounding bias.
\end{abstract}


\section{Introduction}

Understanding the health effects of air pollution is a critically important problem in public health research. There is a vast literature that indicates substantial detrimental impacts of air pollution \citep{cohen2017estimates,burnett2018global,manisalidis2020environmental}. This has helped inform regulatory policy on how to reduce air pollution, and it is important to obtain accurate estimates of the impact of these policies as both the costs and benefits of these policies are in the billions of dollars \citep{portney1990policy}. Due to the importance of this problem and the need for policy-relevant estimates of the impacts of air pollution, there has been a push towards causal inference approaches in environmental epidemiology \citep{dominici2017best, carone2020pursuit, sommer2021assessing}. Causal inference in these settings is difficult for a multitude of reasons, but arguably the biggest impediment to assessing the causal effect of environmental exposures is the potential for unmeasured confounding bias. In the presence of unmeasured confounding, causal effects are not identifiable, and consistent estimation of them is not possible without additional information or assumptions. In some settings, a natural experiment is available that helps to alleviate issues stemming from unmeasured confounding \citep{rich2017accountability}. Examples include pollution reductions due to the Olympic games in Atlanta or Beijing \citep{huang2015ambient} or the decommissioning of power plants leading to large reductions in pollution for nearby areas \citep{luechinger2014air}. Other work has looked to leverage recent advancements in double negative control designs to estimate causal effects in a manner that is robust to unmeasured confounding \citep{schwartz2023effects}. Generally, however, natural experiments or other designs that alleviate concerns about unmeasured confounding are not always available. Given the importance of this problem in air pollution research and its ubiquity in the broader causal inference framework, we develop methods for partial identification of causal effects under unmeasured confounding. We pay particular attention to settings with multiple treatments and outcomes, which we show offer significant advantages for partial identification.

Historically, the most common approach for assessing the unmeasured confounding assumption is sensitivity analysis, which dates back at least as far as \cite{cornfield1959smoking} who argued that there is a causal relationship between smoking and lung cancer. A central goal of sensitivity analysis is to provide bounds on causal estimands that account for potential biases stemming from unmeasured confounding. In simple cases such as with bounded outcomes, general bounds for causal effects can be estimated \citep{manski1990nonparametric}, though typically bounds are estimated based on sensitivity parameters governing the strength of association between the unmeasured confounders and either the treatment or outcome. One approach is to posit a sensitivity parameter for the relationship between the unmeasured variable and treatment, while assuming the worst-case scenario that the unmeasured confounder and outcome are nearly co-linear \citep{rosenbaum1987sensitivity, rosenbaum1988sensitivity, rosenbaum1991sensitivity, rosenbaum2002overt, rosenbaum2002covariance}. Other approaches have two sensitivity parameters that govern the strength of association between the unmeasured variable and both the treatment and outcome. These parameters are intended to be interpretable so that one can reason about potential values for them. This can be done by letting them be parameters of a regression model \citep{rosenbaum1983assessing, mccandless2007bayesian}, parameters dictating relative risks \citep{vanderweele2017sensitivity, ding2016sensitivity}, or partial $R^2$ values between the unmeasured variable and the treatment or outcome \citep{imbens2003sensitivity, small2007sensitivity, veitch2020sense, cinelli2020making, freidling2022sensitivity, chernozhukov2022long}. 

While much of the existing work focuses on a single treatment variable, recent work has examined settings with multiple treatments \citep{wang2017confounder,miao2022identifying, kong2022identifiability, zheng2021copula, zheng2022bayesian}. Under certain assumptions about the nature of confounding, partial $R^2$ values between the unmeasured confounders and treatment can be identified from the observed data. In some cases, such as sparsity in the effects of the treatment on the outcome or null treatment assumptions, the causal effects are even identifiable in this setting \citep{wang2017confounder,miao2022identifying}. Another example is \citet{kong2022identifiability}, which shows identifiability with multiple treatments and a binary outcome under the assumption the treatments are independent given a common unmeasured confounder. Similar results have been shown in the multi-outcome setting, where multiple outcomes or outcome-side factor structure can provide information about unmeasured confounding \citep{zhou2020promises, zheng2023sensitivity}. The most directly related setting is one with both multiple treatments and multiple outcomes. Recent work in \citet{wu2023blessings} show that causal identification can be obtained under certain conditional independence assumptions related to those seen in the proximal causal inference literature \citep{kuroki2014measurement,tchetgen2024introduction}.
\citet{miao2025leveraging} also studies multiple treatments and outcomes and introduces a causal specificity framework. This work assumes upper bounds on the number of direct causal effects across the treatments and outcomes and obtains testing and identification results under sparsity conditions on the breadth of causal effects.
In general, the presence of multiple treatments or outcomes can provide substantial benefits, which is particularly important for air pollution research as there are large numbers of environmental exposures of interest, and they are potentially associated with a variety of adverse health outcomes. As mentioned above, one can obtain identification in these settings even with unmeasured confounding, albeit under fairly strong, restrictive assumptions. Under weaker assumptions, one can potentially obtain partial identification of treatment effects, where only a set of values of the causal effect is identified, which is the focus of this work.

In this paper, we develop a general partial identification framework for scenarios with multiple treatments and multiple outcomes to study whether the estimated health effects of air pollution are robust to the presence of unmeasured confounding bias.
Our framework is complementary to the existing literature but differs in both assumptions and target. Whereas many related approaches use multiple treatments and outcomes to obtain point identification under proxy, negative controls, conditional independence, sparsity or specificity assumptions, we target joint partial identification regions under a factor confounding assumption. In particular, we do not require the treatment effects on outcomes to be sparse, and we do not aim for point identification from latent factor structure. Instead, we quantify the range of causal effects compatible with the observed multivariate dependence under assumptions about unmeasured confounding. This distinction is important in environmental mixture applications, where many pollutants may plausibly affect many health outcomes.
While motivated by settings in air pollution epidemiology, this work has broader implications for partial identification of causal effects with multiple treatments and outcomes. Under a factor confounding assumption, we are able to identify worst-case bounds for causal effects without any additional sensitivity parameters.  Further, we show that additional assumptions about the bias or effect size on any set of estimands can decrease the worst-case bounds for other causal estimands, highlighting the benefit of joint inference over one-at-a-time sensitivity analysis. Specifically, we illustrate two different classes of assumptions which we can be used reduce the joint partial identification region for all causal estimands: 1) assumptions about the magnitude of effect sizes, for which negative controls is a special case, and 2) assumptions about the magnitude of confounding. We provide theoretical results for these strategies and develop novel computational tools for characterizing partial identification regions. We then estimate the health effects of multivariate air pollution exposures in a nationwide study in the United States using Medicare claims data. We utilize the aforementioned approaches to study whether the adverse health effects of air pollution are robust to unmeasured confounding, and quantify how much the partial identification regions for causal effects of interest change under different assumptions about the strength of confounding.

\section{Motivating study of the health effects of air pollution}
\label{sec:DataDescription}

Understanding the health effects of air pollution is a critically important problem in public health research. While there is a large literature showing adverse health effects of air pollution, nearly all of the research is observational, and therefore relies on certain untestable assumptions. A large number of these studies assume that the observed covariates contain all confounders of the exposure-outcome relationship. Due to the importance of this public health problem, and the impact these studies have on environmental regulatory policy, it is important to assess whether these findings are robust to unmeasured confounding bias. This manuscript works to develop methodology to estimate the causal effect of ozone and specific components of PM$_{2.5}$  on a variety of  health outcomes, while accounting for the possibility that these are subject to unmeasured confounding bias. 

To study this question, we use claims data from Medicare Fee-for-Service enrollees over the age of 65 that were enrolled in the years 2000 to 2016 obtained from the Centers of Medicare and Medicaid Services. For each enrollee we have information on their age, sex, race, and whether they are Medicaid eligible, which functions as a proxy for low socioeconomic status, as well as their residential zip code. In addition to these individual level characteristics, we have a number of area level characteristics that are available through the United States Census Bureau and the Center for Disease Control's Behavioral Risk Factor Surveillance System. These include covariates such as average body mass index, education level, population density, smoking rates, median household income and house value, and the percentage of owner occupied housing. We also adjust for meteorological variables, such as temperature and precipitation rates from the National Climatic Data Center. 

In our analysis, we focus on six different air pollution exposures from two distinct publicly available sources: we obtain estimates of ammonium (NH$_4^+$), nitrates (NO$_3^-$), and sulfate (SO$_4^{2-}$) levels on a ($0.01^\circ \times 0.01^\circ$) grid from the Atmospheric Composition Analysis Group \citep{van2019regional} and we obtain estimates of elemental carbon (EC), organic carbon (OC), and ozone (O$_3$) on a 1km by 1km daily grid from the Socioeconomic Data and Applications Center \citep{di2021daily, requia2021daily}. These exposures are aggregated to the zip code level so that they can be linked to the residential zip code obtained from Medicare, and are additionally aggregated to the yearly level as we are examining long-term exposure to pollution.  With the exception of ozone, the exposures considered in this study are decreasing on average across the United States over the time frame considered, though there exists variability in this process across zip codes.  The outcomes of interest consist of zip code level hospitalization rates for anemia, COPD, lung cancer, stroke, and asthma, which are obtained based on ICD9/10 diagnosis codes. Additionally, we have access to negative control outcomes, which are hospitalization rates in the year prior to exposure. We focus on using hospitalization rates for COPD in the prior year as a negative control outcome. Note that an alternative approach would be to adjust for prior outcomes such as these as observed covariates, though we show in subsequent sections that they can be particularly useful within our framework for sharpening inferences about causal effects of interest when they are treated as negative control outcomes. Our unit of analysis is a United States zip code, and our data consists of all zip codes in the contiguous United States. All covariates are aggregated up to the zip code level by taking their averages or proportions within each zip code. Similar data sources have been used previously to provide impactful, large-scale estimates of the health effects of PM$_{2.5}$ \citep{di2017air}. Critically, these results have relied on the no unmeasured confounding assumption. While we have measured a wide range of demographic, meteorological, and health characteristics for the zip codes examined, it is very possible that these are insufficient to fully address confounding. For this reason, and due to the importance of this public health issue, in this work we aim to clarify the extent to which these effects are robust to such unmeasured confounding. This will provide environmental regulators with improved scientific evidence on the nature of the relationship between air pollution and public health. 

\section{Notation and estimands}
Throughout, we observe $n$ independent and identically distributed observations of $(Y, T, X)$, where $T$ denotes a vector of $k$ treatments, $Y$ denotes a vector of $q$ outcomes, and $X$ denotes a vector of $l$ pre-treatment covariates, respectively. Throughout, we use the terms exposures and treatments interchangeably when referring to $T$. Under the potential outcome framework \citep{rubin1974estimating, splawa1990application}, we let $Y(t)$ denote the potential outcome that would have been observed had the treatment $T$ been set to $t$. Our goal throughout will be to utilize the observed data to estimate the population average treatment effect (PATE), for a linear combination of outcomes $a^{\top}Y$, defined as
\begin{equation*}
    \text{PATE}_{a,t_1,t_2} := E[a^\top Y(t_1) - a^\top Y(t_2)]
\end{equation*}
for some $t_1, t_2$. Note that throughout we use subscripts for treatment vectors to denote distinct treatment vectors, and these do not refer to specific components of the vector $t$. If unmeasured confounders $U$ were observed along with $X$, we could nonparametrically identify the causal effect from the observed data under the following assumption:
\begin{assumption}[] \label{ass1-3}
\hfill
\begin{enumerate}
    \item SUTVA (Stable Unit Treatment Value Assumption): There is no interference between units and there is only a single version of each assigned treatment.
    \item Latent ignorability: $Y(t) \independent T \mid (U, X)$ for all $t$.
    \item Latent positivity: $f(t \mid U=u, X=x) > 0$ for all possible values of $t$, $x$, and $u$. Here, $f(\cdot)$ represents the conditional density function of $T$ given $(X, U)$.  
\end{enumerate}
\end{assumption}

The SUTVA assumption ensures that $Y(t)$ is well-defined and that $Y=Y(t)$ if $T=t$, which links the observed data to potential outcomes. The latent ignorability assumption states that $X$ and $U$ contain all common causes of $T$ and $Y$. The latent positivity assumption, also referred to as overlap, assures that there is a positive probability that each unit receives any treatment value $t$ given both $X$ and $U$. Under these assumptions, we can identify the causal effect of interest using $E[Y(t)] = E_{X,U}[E(Y|T=t,X,U)]$. 

Note that our identification assumptions include unmeasured confounders $U$, which are not observed, and therefore the causal effect is not identifiable without further assumptions, as the effect of $T$ on $Y$ may be confounded by $U$ even after conditioning on $X$. Our goal in this manuscript is to derive partial identification regions for the causal effect that account for the potential presence of unmeasured confounders when we have both multiple treatments and multiple outcomes under varying assumptions about the unmeasured variables. For now, we assume the number of unmeasured confounders $m$ is known and fixed. We discuss estimation of $m$ in Appendix \ref{sec:AppendixEstimationFactorModel}.

\section{Partial Identification and Factor Confounding} \label{sec-method}

In this section, we first show that the causal effect of interest is not point-identified in our setting, but the range of possible values for the causal effect can be characterized by quantities that, under a factor confounding assumption, can be identified in the multi-treatment and multi-outcome setting. Note that throughout this section we omit intercepts from all models for ease of exposition. Additionally, without loss of generality, we assume that $E(U) = 0$. 

\subsection{Factor models with multivariate treatments and outcomes}
For ease of exposition, we do not include observed covariates $X$ in this section, though all results hold analogously when conditioning on $X$. Throughout, we consider the latent variable models that are described by the following assumptions.   

\begin{assumption}[Confounder homoskedasticity]
\label{asm:gaussian_latent}
\sloppy
Let $E[U|T=t] := \mu_{u|t}$ and $Var(U|T=t) := \Sigma_{u|t}$ be the $m$-dimensional mean and $m \times m$ covariance matrix of latent confounders given treatments. Assume that the confounders are homoskedastic in $t$, meaning that $\Sigma_{u|t}$ is invariant to the level of $t$.  
\label{confounder_mean_var}
\end{assumption}

\begin{assumption}[Outcomes are linear in latent confounders]
The mean of the outcome is linear in the unmeasured confounders with mean
\begin{align}
    E[Y|T, U] &= g(T) + \Gamma \Sigma_{u|t}^{-1/2} U.~~~ \label{Y}
\end{align} with  $\Gamma \in \mathbb{R}^{q \times m}$. We further assume a $q$-dimensional diagonal variance $Var(Y|T, U) := D$.
\label{outcome_mean}
\end{assumption}
\noindent Note that if we had observed covariates $X$, they would be placed inside of the $g(\cdot)$ function along with $T$ thereby allowing for arbitrary relationships between the observed variables. We only require a specific structure for the effect of the unobserved confounders on the outcome. 

Assumptions \ref{confounder_mean_var} and \ref{outcome_mean} ensure that the conditional variance of $U$ given exposures does not depend on the level of exposure, and that the effect of $U$ on the outcome does not depend on the exposures. Together, these assumptions are needed in order to obtain a closed-form expression for the bias of the causal effect that depends only on parameters that can be partially identified under an additional factor confounding assumption that follows. Note that the combination of these two assumptions is still significantly weaker than assuming no unmeasured confounding, and allows for a broad class of unmeasured confounding mechanisms. Under these assumptions, the observed outcome moments are
\begin{align}
    E(Y | T=t) &:= \check{g}(t) = E[E(Y | T=t, U)] = g(t) + \Gamma \Sigma_{u|t}^{-1/2} \mu_{u|t}, \label{eqn:OutcomeMeanBias}\\
    \text{Var}(Y | T=t) &= \text{Var}[E(Y | T=t, U)] + E[\text{Var}(Y | T=t, U)] =
    \Gamma \Gamma^{\top} + D. \label{AR56}
\end{align}
Given that our estimand of interest is
\begin{align}
    \text{PATE}_{a,t_1,t_2} &:= E[a^{\top}Y(t_1)-a^{\top}Y(t_2)] \\
    &= a^{\top}g(t_1)-a^{\top}g(t_2),
\end{align} 
Equation \ref{eqn:OutcomeMeanBias} shows that that the bias from ignoring unmeasured confounders, $U$ is
\begin{align} \label{bias}
    \text{Bias}_{a,t_1,t_2}
    &= [a^{\top}\check{g}(t_1)-a^{\top}\check{g}(t_2)] - \text{PATE}_{a,t_1,t_2} \nonumber\\
    &= a^{\top} \Gamma \Sigma_{u|t}^{-1/2}(\mu_{u|t_1}-\mu_{u|t_2}).
\end{align} 

While the bias term in equation \eqref{bias} is not identifiable in the absence of data on $U$, we can derive upper bounds for this bias that are themselves identifiable under certain assumptions about the unmeasured confounders. Note that moving forward, we define $\mu_{u|\Delta t} = \mu_{u|t_1} - \mu_{u|t_2}$.
\begin{theorem} \label{thm-biasbound}
    Given Assumptions \ref{ass1-3}-\ref{outcome_mean}, the confounding bias of $\text{PATE}_{a,t_1,t_2}$ can be bounded as
    \begin{equation*}
        \text{Bias}_{a,t_1,t_2}^2 \leq  \|a^{\top}\Gamma\|_2^2 \| \Sigma_{u|t}^{-1/2} \mu_{u|\Delta t} \|^2_2,
    \end{equation*}
    where the right-hand side is the worst-case bias of the naive estimator and is achieved when $a^{\top}\Gamma$ is colinear with $\Sigma_{u|t}^{-1/2} \mu_{u|\Delta t}$.
\end{theorem}
The proof of the theorem is given in Appendix \ref{sec:techdetails}. $a^{\top}\Gamma$ can be viewed as the strength of association between the unmeasured variables $U$ and the outcome, while $\Sigma_{u|t}^{-1/2} \mu_{u|\Delta t}$ is the scaled difference in unmeasured confounder means when treatment is shifted from $t_1$ to $t_2$ and represents the strength of association between the unmeasured confounders and treatment. This result implies that the true treatment effect lies in the interval 
\begin{equation}\label{interval}
    a^{\top}[\check{g}(t_1)-\check{g}(t_2)] \pm \|a^{\top}\Gamma\|_2 \| \Sigma_{u|t}^{-1/2} \mu_{u|\Delta t} \|_2.
\end{equation}
 While the confounding bias itself is not identifiable, it can be shown that this bound is identified for some latent variable models. In particular, in this work we focus on the probabilistic PCA model \citep{tipping_and_bishop} to model covariation among treatments:
\begin{align}
U &= \epsilon_U \label{eqn:u}\\
T &= BU + \epsilon_T \label{eqn:t}
\end{align}
\noindent where $B \in \mathbb{R}^{k \times m}$, $\epsilon_U \sim N_m(0,I_m)$ and $\epsilon_{T} \sim N_k(0,\Sigma_{t|u})$ with $\Sigma_{t\mid u}$ being a diagonal matrix. This model implies that the treatment levels are a linear function of a small number of unobserved latent factors plus idiosyncratic Gaussian error. Note that a key consequence of our model formulation, and the fact that both $D$ and $\Sigma_{t|u}$ are diagonal matrices, is that we assume that all residual correlation in the exposures and outcomes is due to the unmeasured variables $U$. Under this model the conditional confounder distribution is
\begin{equation} \label{f_u|t}
    f(u|T=t) \sim N_m(\mu_{u|t}, \Sigma_{u|t})
\end{equation}
where
\begin{align*}
    \mu_{u|t}
    &= B^{\top}(BB^{\top}+\Sigma_{t|u})^{-1}t, \\
    \Sigma_{u|t}
    &= I_m - B^{\top}(BB^{\top}+\Sigma_{t|u})^{-1}B.
\end{align*}

This leads to our main assumption, which will allow us to identify bounds on the causal effect.

\begin{assumption}[Factor confounding assumption]
Assume $rank(\Gamma) = m_1$ and $rank(B) = m_2$ with $m = m_1 + m_2$.  Without loss of generality define $\Gamma = \tilde \Gamma R$ where $R \in O(m)$ is an $m \times m$ orthonormal matrix such that $\tilde \Gamma \tilde \Gamma^{\top } = \Gamma \Gamma^{\top }$ with
\begin{align}
\tilde \Gamma &= \left[\begin{array}{ll}\Gamma_1 & \mathbf{0}_{q \times m_2}\end{array}\right] \\
B &=\left[\begin{array}{ll}
\mathbf{0}_{k \times m_1} & B_1
\end{array}\right],
\end{align}
where $\Gamma_1 \in \mathbb{R}^{q \times m_1}$ and $B_1 \in \mathbb{R}^{k \times m_2}$. 
\label{ass-iden4}
Further, assume that the treatments are generated by \eqref{eqn:u}--\eqref{eqn:t} and $U$ represents potential confounders in the sense that they are possible causes of $T$ and $Y$ and are not themselves caused by $T$ or $Y$. Additionally, we assume the following conditions are satisfied.
    \begin{enumerate}
        \item\label{ass4-iden-C1} The number of outcomes and unmeasured confounders satisfies $(q-m_1)^2-q-m_1 \geq 0$, and if any row of $\Gamma_1$ is deleted, there remain two disjoint submatrices of rank $m_1$, where $m_1=\text{rank}(\Gamma_1)$
        \item\label{ass4-iden-C2} The number of treatments and unmeasured confounders satisfies $(k-m_2)^2-k-m_2 \geq 0$, and if any row of $B_1$ is deleted, there remain two disjoint submatrices of rank $m_2$, where $m_2=\text{rank}(B_1)$.
    \end{enumerate}
\end{assumption}

The zero-padded construction for $\tilde \Gamma$ and $B$ fixes an initial reference parameterization in an $m$-dimensional latent space. Under an arbitrary orthogonal rotation $R \in O(m)$, the rotated outcome loading matrix $\Gamma = \tilde{\Gamma} R$ allows for any degree of alignment (confounding) between treatment and outcome variations while preserving the marginal residual covariances, $\Gamma\Gamma^{\top}$. This is formalized subsequently in Theorem \ref{thm:bias_identifiability}. Additionally note that this allows for the variables in $U$ to be confounders, or solely predictors of the treatment or outcome, depending on the orthogonal rotation matrix $R$.

Assumptions \ref{ass-iden4}.1 / \ref{ass-iden4}.2 are standard in the factor modeling literature \citep{anderson1956statistical}. When violated, $\Gamma$ and $B$ are not identifiable even up to rotation, as it becomes impossible to separate residual variation in the exposures or outcomes that is due to the potential unmeasured confounders $U$, from  the diagonal idiosyncratic residual variation. While the conditions regarding the submatrix of $B$ and $\Gamma$ are somewhat abstract, they have important real-world applications that should be reasoned through before making such an assumption. Specifically, these assumptions require that each unmeasured confounder be associated with at least three exposures or three outcomes. This type of shared confounding structure is something that should be considered carefully before making the factor confounding assumption. We discuss the plausibility of factor confounding for our air pollution study in detail in Section \ref{ssec:MedicareAssumptions}. Note that if factor confounding is only plausible for the exposures or for the outcomes, but not both, then one can move forward with only the plausible factor confounding assumption. In this case, however, one must also posit sensitivity parameters that determine the strength of association between the unmeasured variables and either the exposures or outcomes, whichever does not have the factor confounding assumption applied to it. We focus on settings where this assumption is made for both exposures and outcomes throughout the manuscript, but we provide details on how inference would proceed in this situation, along with an application to the Medicare data, in Appendix \ref{sec:AppendixExtraMedicare}. Importantly, when the factor confounding assumption does hold for both exposures and outcomes, then bounds on the unmeasured confounding bias are identifiable.

\begin{theorem}
Given Assumptions \ref{ass1-3}--\ref{ass-iden4}, both $\|a^{\top}\Gamma\|_2^2$ and $\| \Sigma_{u|t}^{-1/2} \mu_{u|\Delta t} \|^2_2$ are identifiable and thus the bound on $\text{Bias}_{a,t_1,t_2}$, given by $\|a^{\top}\Gamma\|_2^2 \| \Sigma_{u|t}^{-1/2} \mu_{u|\Delta t} \|^2_2$, is also identifiable. There exists an $m \times m$ orthonormal matrix, $R \in O(m)$, such that
$$\text{Bias}_{a,t_1,t_2} =  a^{\top} \widetilde \Gamma R  \Sigma_{u|t}^{-1/2} \mu_{u|\Delta t}.$$   $\text{Bias}_{a,t_1,t_2}$ is identifiable $\forall ~ a, t_1, t_2$ if and only if $R$ is known. Further, $\text{Bias}_{a,t_1,t_2} = 0 ~  \forall ~ a, t_1, t_2 $ if and only if $R$ is block diagonal, that is, it can be written as $$R = \begin{pmatrix}
R_{11} & \mathbf{0}_{m_1 \times m_2} \\
\mathbf{0}_{m_2 \times m_1} & R_{22}
\end{pmatrix}.$$
\label{thm:bias_identifiability}
\end{theorem}
In addition to showing that the bounds are identifiable, Theorem \ref{thm:bias_identifiability} also shows that even though both $\Gamma$ and $B$ are only identifiable up to rotation, the bias can be written as a function of a single rotation matrix $R$. This allows us to effectively fix $B$ and explore different biases by exploring different rotations of the outcome parameters $\Gamma$.

\section{Joint Partial identification regions under additional constraints} \label{ssec:MultipleEstimands}

\begin{figure}[h]
    \centering
    \includegraphics[width=1\linewidth]{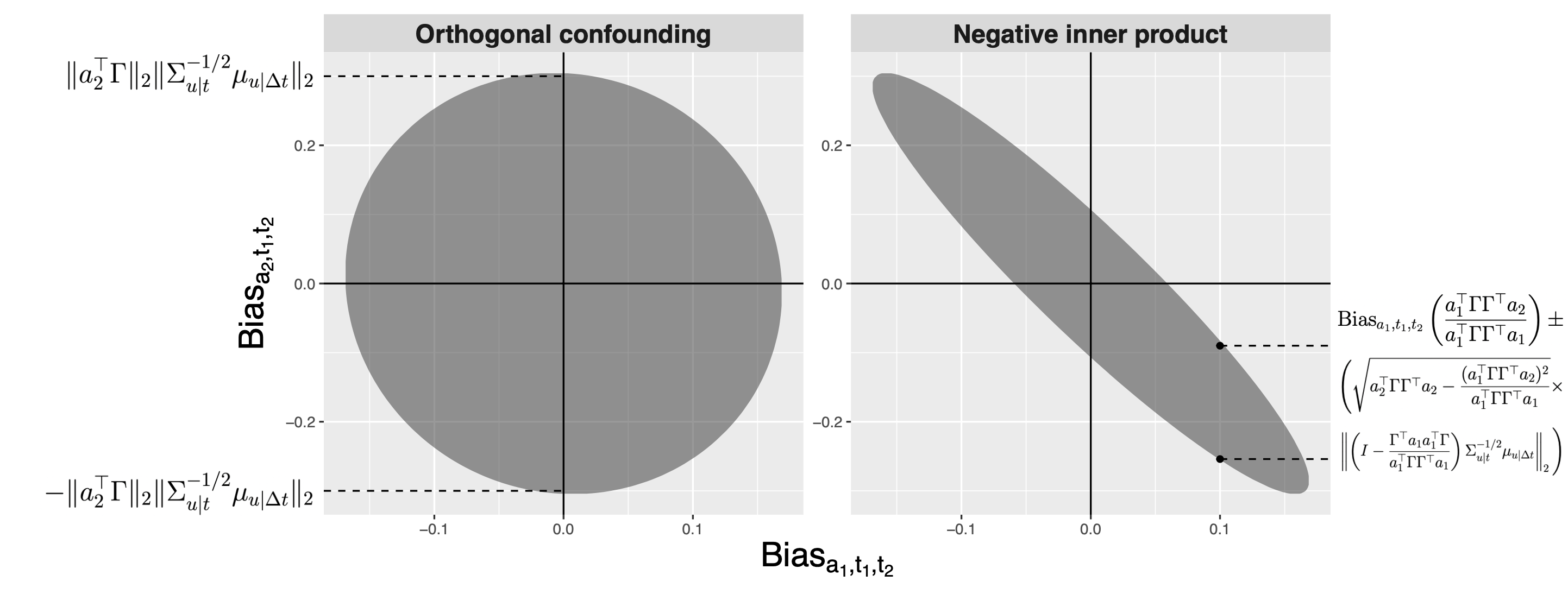}
    \caption{Comparison of bivariate partial identification regions for the bias when either $\langle a_1^{\top} \Gamma, a_2^{\top} \Gamma \rangle = 0$ or $\langle a_1^{\top} \Gamma, a_2^{\top} \Gamma \rangle < 0$. The horizontal dashed lines in the left panel correspond to the marginal partial identification region for the bias of the second estimand. The horizontal dashed lines in the right panel correspond to the conditional partial identification region for the bias of the second estimand for a particular choice of the bias of the first estimand. }
    \label{Fig:BivariateIllustration}
\end{figure}

Theorem \ref{thm:bias_identifiability} highlights that with a factor confounding assumption, we can bound the causal effects for all estimands in \eqref{interval} in settings with multiple treatments and multiple outcomes. However, in some contexts, these bounds are too wide to be practically useful without additional assumptions. This occurs in our motivating example in Section \ref{sec:Application} as the bounds contain all plausible (and some implausible) effects of air pollution. However there are additional benefits that come from considering the joint partial identification region for multiple estimands. Importantly, the joint partial identification region is \emph{not} simply the Cartesian product of the marginal bounds, as  the worst-case bounds for multiple estimands cannot be simultaneously attained. In many instances, assumptions about the bias on one estimand can be highly informative about the bias of other estimands (see Figure \ref{Fig:BivariateIllustration}).

By way of example, suppose that we are interested in two estimands, $\text{PATE}_{a_1, t_1, t_2}$ and $\text{PATE}_{a_2, t_1, t_2}$, which correspond to distinct outcomes, $a_1$ and $a_2$, but the same treatment contrast ($t_2$ vs $t_1$).   
The degree to which assumptions about $\text{PATE}_{a_1, t_1, t_2}$ inform the partial identification region for $\text{PATE}_{a_2, t_1, t_2}$ is driven in part by the inner product of $a_1^{\top} \Gamma$ and $a_2^{\top} \Gamma$. This is illustrated in Figure \ref{Fig:BivariateIllustration} where we can see a strong dependence in the bivariate partial identification region when $|\langle a_1^{\top} \Gamma$, $a_2^{\top} \Gamma \rangle|$  is large (Figure \ref{Fig:BivariateIllustration}, right). Even when $a_1^{\top} \Gamma$ and $a_2^{\top} \Gamma$ are orthogonal,  the magnitude of the bias for one estimand is weakly informative about the magnitude of the bias for the other (Figure \ref{Fig:BivariateIllustration}, left). We formalize these ideas in the following proposition, which characterizes the partial identification region for one estimand as a function of the bias of another. 

\begin{proposition}
\label{prp:bias_innerproduct}
Given Assumptions \ref{ass1-3}--\ref{ass-iden4}, if we further assume that $\text{Bias}_{a_1,t_1,t_2} = \beta_1$,
then the partial identification region for a different outcome with the same treatment contrast, $\text{Bias}_{a_2,t_1,t_2}$, is given by
\begin{align}
    \text{Bias}_{a_2,t_1,t_2} &\in \nonumber \\
    \beta_1\left( \frac{a_1^{\top} \Gamma \Gamma^{\top} a_2}{a_1^{\top} \Gamma \Gamma^{\top} a_1} \right) 
&\pm \sqrt{a_2^{\top} \Gamma \Gamma^{\top} a_2 - \frac{(a_1^{\top} \Gamma \Gamma^{\top} a_2)^2}{a_1^{\top} \Gamma \Gamma^{\top} a_1}}  \left\| \left(I - \frac{\Gamma^{\top} a_1 a_1^{\top} \Gamma}{a_1^{\top} \Gamma \Gamma^{\top} a_1} \right) \Sigma_{u|t}^{-1/2} \mu_{u|\Delta t} \right\|_2 \label{conditional_region}\\
    &\subseteq \left[ -\|a_2^{\top}\Gamma\|_2 \| \Sigma_{u|t}^{-1/2} \mu_{u|\Delta t} \|_2, ~~~ \|a_2^{\top}\Gamma\|_2 \| \Sigma_{u|t}^{-1/2} \mu_{u|\Delta t} \|_2 \right] \label{marginal_region}
\end{align}
\end{proposition}
This conditional partial identification region \eqref{conditional_region} is contained in the marginal identification region \eqref{marginal_region}, can be much shorter in length, and may have a different center. For one, given $\text{Bias}_{a_1,t_1,t_2} = \beta_1$, the identification region for $\text{Bias}_{a_2,t_1,t_2}$  is no longer necessarily centered at zero. The magnitude of the bias correction depends on $\beta_1$ as well as  $\langle a_1^{\top} \Gamma, a_2^{\top} \Gamma \rangle$. The width of the partial identification region in \eqref{conditional_region} is the product of two terms. The first term is smallest when $|\langle a_1^{\top} \Gamma, a_2^{\top} \Gamma\rangle|$ is large and reaches zero when $a_1^{\top}\Gamma$ and $a_2^{\top}\Gamma$ are collinear. In the air pollution setting, 
it is plausible that 
 $|\langle a_1^{\top} \Gamma, a_2^{\top} \Gamma\rangle |$ is large, as common unmeasured confounders are likely to affect multiple public health outcomes in similar ways. The second term is small when $\text{Bias}_{a_1,t_1,t_2}$ is large in magnitude. This is because $\bigg(I - \frac{\Gamma^{\top} a_1 a_1^{\top} \Gamma}{a_1^{\top} \Gamma \Gamma^{\top} a_1} \bigg) \Sigma_{u|t}^{-1/2} \mu_{u|\Delta t}$ is the projection of $\Sigma_{u|t}^{-1/2} \mu_{u|\Delta t}$ onto the space orthogonal to $a_1^{\top} \Gamma$, which is smaller when the bias, $a_1^{\top} \Gamma \Sigma_{u|t}^{-1/2} \mu_{u|\Delta t}$, is large. 

While we have focused on two estimands with the same exposure contrast and different outcomes, similar intuition holds for two estimands with the same outcome and different exposure contrasts. If two estimands have different treatment contrasts, then they will have different values of the $\Sigma_{u|t}^{-1/2} \mu_{u|\Delta t}$ term in their respective biases, and the inner product between these vectors dictates the strength of dependence in the partial identification region. We don't explicitly write down these partial identification regions, as they can be seen as special cases of the partial identification regions we derive later in Section \ref{ssec:NegControls} that hold in more general settings for any number of estimands.  

To take full advantage of the benefits of characterizing the joint partial identification region, we can make additional useful assumptions that can be used to reduce the partial identification regions for all estimands.  In the following, we discuss several different kinds of such assumptions.

\subsection{Sensitivity Bounds on treatment effects}
\label{ssec:BoundsTreatmentEffects}

In many applications, prior knowledge on a reasonable range of effect sizes is known and can be incorporated into the partial identification regions. For instance, in our application to air pollution (Section \ref{sec:Application}), we know that it is biologically implausible for pollution to have strong protective effects on health outcomes. We may also have strong prior knowledge about the largest plausible harmful effect, as it is not expected for air pollution to be the sole driver of complex public health outcomes.  In our notation, a direct assumption about the magnitude of the causal effect on a single estimand takes the form
$$z_l \leq E[a^{\top}Y(t_1)-a^{\top}Y(t_2)] \leq z_u, $$
for a given choice of upper and lower bounds on the treatment effect, $z_u$ and $z_l$. These bounds provide information on the rotation matrix, $R$, via the following inequality:
\begin{align}
    z_l \leq [a^{\top}\check{g}(t_1)-a^{\top}\check{g}(t_2)] - a^{\top} \widetilde \Gamma R \Sigma_{u|t}^{-1/2} \mu_{u|\Delta t} \leq z_u. \label{eqn:SoftConstraint}
\end{align}
If we make these assumptions for a number of estimands, we gain more information about $R$, and hence the bias for all other estimands. Analytic characterizations of the partial identification region under these constraints is a non-trivial problem and as such, we find approximate solutions using a rejection sampling algorithm to explore the range of possible causal effects satisfying \eqref{eqn:SoftConstraint}. Specifically, we draw orthogonal matrices $R^{(h)}$ for $h = 1, \dots, H$ uniformly on the Stiefel manifold of $m \times m$ orthogonal matrices, for large values of $H$. For each orthogonal matrix, $R^{(h)}$, we check whether the constraints in \eqref{eqn:SoftConstraint} are satisfied. Each matrix $R^{(h)}$ that satisfies the aforementioned criteria determines the bias for each estimand of interest, and we can explore the set of these matrices to determine plausible biases, and therefore partial identification regions. Since $m$ is generally small (e.g. single a digit value), such a Monte Carlo rejection sampling strategy is feasible.  In Appendix  \ref{sec:Stiefel}, we provide theoretical support showing that the difference between the true partial identification region and the approximate one based on sampling approaches zero as $H$ grows, and also show empirically in simulation that this difference becomes effectively zero with only moderate values of $H$. 

\subsection{Sensitivity Bounds on Partial R-squared Values}
\label{ssec:PartialRsquaredBounds}

As an alternative to specifying sensitivity bounds directly on the treatment effects, we can instead specify bounds on the partial variance explained by confounders.  In a single-treatment single-outcome analysis, \citet{cinelli2020making} propose specifying upper bounds on both the partial fraction of outcome variance explained by confounders ($R^2_{Y\sim U | T}$) and the fraction of treatment variance explained by confounders ($R^2_{T \sim U})$. Here, we show how similar approaches incorporating bounds on the confounding strength can be incorporated into our multi-treatment multi-outcome setting. Throughout this section, we let $d = (t_1 - t_2)$ correspond to the treatment contrast of interest. For a given treatment-outcome pair, $a^\top Y$ and $d^\top T$, let $U_{a,d}$ represent a scalar projection of $U$ which captures all confounding for that treatment-outcome pair.  We call $U_{a,d}$ a ``minimal deconfounder" for treatment-outcome pair $a^\top Y$ and $d^\top T$ since it is the lowest dimensional function of $U$ that is sufficient to address confounding \citep{d2021deconfounding}. There is no unique minimal deconfounder: for example, both $d^{\top}BU$ and $a^{\top}\Gamma\Sigma_{u\mid t}^{-1/2} U$ are different scalar projections of $U$ that are sufficient to remove all confounding bias. We move forward with $U_{a,d} = a^{\top}\Gamma\Sigma_{u\mid t}^{-1/2} U$ as it makes certain calculations more tractable and is a minimal deconfounder for all exposure contrasts for the choice of outcome given by $a$. 

This choice of minimal deconfounder essentially maximizes the association between $U_{a,d}$ and $Y$ that is allowed under the factor confounding assumption, and one can show that the partial $R^2$ between this confounder and the treatment contrast, given covariates, is given by
$$R^2_{d^{\top}T \sim U_{a,d}}  = \frac{\left( a^{\top}\Gamma\Sigma_{u\mid t}^{-1/2} B^\top d \right)^2 }{|| a^{\top}\Gamma\Sigma_{u\mid t}^{-1/2} ||_2^2 \left( d^\top (B B^\top + \Sigma_{t|u}) d \right) }.$$
This quantity is a function of $B$ and $\Gamma$, and therefore bounds on $R^2_{d^{\top}T \sim U_{a,d}}$ implicitly constrain the rotation matrix $R$, which can translate to reductions in the partial identification region for the causal effects of interest.

One can either reason about these values using subject matter expertise or they can use benchmark values from observed covariates to reason about their magnitude. Finally, as with bounding the treatment effects directly, joint R-squared style sensitivity analysis for multiple estimands can have large benefits relative to a one-at-a-time sensitivity analysis. The partial R-squared values for each estimand are not variationally independent, as they depend on the common rotation matrix parameter $R$, and thus if we require marginal bounds to be satisfied concurrently for all estimands, then the implied joint bounds for individual estimands can be much tighter. To find the partial identification regions, we can use the same sampling algorithm discussed in Section \ref{ssec:BoundsTreatmentEffects}, where we retain rotation matrices that satisfy all of the partial R-squared constraints.

\subsection{Negative controls}
\label{ssec:NegControls}

Negative controls have been widely used in observational studies to detect or mitigate bias in the causal effect of the treatment on the outcome. In air pollution epidemiology, examples include future air pollutant levels or air pollutant levels in a distant area being used as negative control exposures \citep{lumley2000assessing}, and appendicitis hospitalizations being used as a negative control outcome \citep{sheppard1999effects}. Generally, in the multi-treatment, multi-outcome setting, we refer to a negative control pair as any treatment and outcome pair among our $k$ treatments and $q$ outcomes for which we assume \emph{a priori} that the PATE for that treatment-outcome pair is zero. Such negative control assumptions can be viewed as a special case of a sensitivity bound on the treatment effects described in Section \ref{ssec:BoundsTreatmentEffects},  where the constraints in \eqref{eqn:SoftConstraint} are set so that $z_l = z_u = 0.$ Given the importance of null controls in the literature, we elaborate on this special case below.

Suppose throughout that we have $J$ outcomes with at least one treatment known to not causally affect that outcome, and that the $j$th outcome has $c_j \leq m$ such negative control pairs. The $j$th outcome is denoted by $b_j^{\top} Y$, and the treatment contrasts for outcome $j$ are given by $$\left\{ \left(t_1^{(j,1)},t_2^{(j,1)}\right),\cdots,\left(t_1^{(j,c_j)},t_2^{(j,c_j)}\right) \right\}.$$ We further define the difference in the $\check{g}(\cdot)$ function at these contrasts as 
\begin{equation} \label{gcheck}
    \check{\mathcal{G}}_j=\left[\check{g}\left(t_1^{(j,1)}\right)-\check{g}\left(t_2^{(j,1)}\right),\cdots,\check{g}\left(t_1^{(j,c_j)}\right)-\check{g}\left(t_2^{(j,c_j)}\right)\right].
\end{equation}
Negative controls provide information on the rotation matrix $R$ through the following equation:
\begin{equation} \label{NCcondition}
    b_j^{\top} \widetilde{\Gamma} R M_j = b_j^{\top}\check{\mathcal{G}}_j,
\end{equation}
where
\begin{equation} \label{M}
    M_j = \Sigma_{u|t}^{-1/2} \left[\mu_{u|t_1^{(j,1)}} - \mu_{u|t_2^{(j,1)}},\cdots,\mu_{u|t_1^{(j,c_j)}} - \mu_{u|t_2^{(j,c_j)}} \right].
\end{equation}
The left side of \eqref{NCcondition} is the functional form for the bias of negative control estimands, while the right side corresponds to the estimated causal effects under ``no unobserved confounding''. For null controls, these are equal by definition since the true causal estimand is assumed to be zero a priori. This equation restricts the space of possible $R$ matrices, and therefore can reduce the size of the partial identification region for all estimands of interest. 

The partial identification regions with null control constraints can be complex, and therefore we introduce a constrained optimization problem for deriving these partial identification regions under arbitrary negative control constraints. Additionally, we derive analytic partial identification regions that are conservative at times, but provide intuition about the benefits of negative controls. 

\subsubsection{Approaches to finding negative control partial identification regions} \label{sec-NC-numerical} 

For any feasible omitted variable bias, $\beta$, with $|\beta| \leq \|a^{\top}\Gamma\|_2 \| \Sigma_{u|t}^{-1/2} \mu_{u|\Delta t} \|_2$, to be consistent with the negative control assumption in \eqref{NCcondition}, there must be an orthogonal matrix $R$ such that the following conditions are satisfied:
\begin{enumerate}
    \item $a^{\top}\widetilde{\Gamma} R \Sigma_{u|t}^{-1/2} \mu_{u|\Delta t} = \beta $.
    \item  $b_j^{\top}\widetilde{\Gamma} R M_{j} = b_j^{\top}\check{\mathcal{G}}_j$ for all $j = 1, \dots, J$ negative control contrasts.
\end{enumerate}
The first condition establishes that for rotation matrix $R$ the omitted variable bias for outcome $a^{\top}Y$ is indeed $\beta$.  The other conditions ensure that the orthogonal matrix $R$ also implies the correct omitted variable biases for the negative controls. For certain values of $\beta$, there may not exist a matrix $R$ that satisfies the null control conditions. We can express the partial identification region for $ \text{Bias}_{a,t_1,t_2}$ as the set  $\mathcal{B}_{a, t_1, t_2} = \left\{ \beta : \nu_\beta = 0 \right\}$ where
\begin{align}
\nu_\beta := \min_{\widetilde{R} \in \mathcal{V}_{m,m}}  \left((a^{\top} \widetilde{\Gamma} \widetilde{R} \Sigma_{u|t}^{-1/2} \mu_{u|\Delta t} - \beta)^2 + \sum_{j=1}^J \| b_j^{\top} \widetilde{\Gamma} \widetilde{R} M_{j} - b_j^{\top}\check{\mathcal{G}}_j \|_2^2\right),\label{eqn:stiefel}
\end{align}
and $\mathcal{V}_{m,m}$ is the Stiefel manifold of all $m \times m$ orthogonal matrices. If the bias value $\beta$ is in the partial identification region established by the negative control constraints, then $\nu_\beta = 0$. Therefore, for any $\beta$, we minimize equation \eqref{eqn:stiefel} over the Stiefel manifold using the optimization algorithm implemented in the \texttt{rstiefel} package \citep{rstiefel}.  To find the partial identification region, we minimize  \eqref{eqn:stiefel} over a grid of all possible values of $\beta$ and keep track of those that minimize this objective function at zero. In practice, the numerical solutions $\nu_\beta$ are never exactly zero, and thus we simply check that both $(a_1^{\top} \widetilde{\Gamma} \widetilde{R} \Sigma_{u|t}^{-1/2} \mu_{u|\Delta t} - \beta)^2 \leq \delta_1$ and $\sum_{j=1}^J \| b_j^{\top} \widetilde{\Gamma} \widetilde{R} M_{j} - b_j^{\top}\check{\mathcal{G}}_j \|_2^2 \leq \delta_2$ for sufficiently small constants $\delta_1, \delta_2 > 0$. A full description of the algorithm for identifying feasible biases can be found in Appendix \ref{app-sec-NC-numerical}. 

In addition to finding the bounds computationally, we are able to derive partial identification regions analytically. While these regions can be conservative in certain situations, they provide useful insight into the types of negative controls that lead to the largest reduction in the size of the partial identification region.  

\begin{theorem} \label{thm-NCsingle}
Suppose we only have $J=1$ negative control outcome with $c_1 \leq m$ negative control contrasts and that Assumptions \ref{ass1-3}--\ref{ass-iden4} hold. Let $k_{aa} = a^{\top} \Gamma \Gamma^{\top} a$, $k_{ab} = a^{\top} \Gamma \Gamma^{\top} b_1$, and $k_{bb} = b_1^{\top} \Gamma \Gamma^{\top} b_1$, which are all identifiable scalars from our factor model assumptions. Lastly, let $M_1^{\dagger}$ be a generalized inverse of $M_1$. Then, $\text{Bias}_{a,t_1,t_2}$ is in the following interval
\begin{align} \label{NCinterval-single}
    & \frac{k_{ab}}{k_{bb}} b_1^{\top}\check{\mathcal{G}}_1 M_{1}^{\dagger} \Sigma_{u|t}^{-1/2} \mu_{u|\Delta t} \quad \pm \nonumber \\
    & \Bigg( \Big| \frac{k_{ab}}{k_{bb}} \Big| \sqrt{k_{bb} - \| b_1^{\top}\check{\mathcal{G}}_1 M_1^{\dagger} \|_2^2} \| (I - M_1 M_1^{\dagger}) \Sigma_{u|t}^{-1/2} \mu_{u|\Delta t} \|_2 
    + \sqrt{k_{aa} - \frac{k_{ab}^2}{k_{bb}}} \ \|\Sigma_{u|t}^{-1/2} \mu_{u|\Delta t} \|_2\Bigg),
\end{align}
\end{theorem}

The proof of Theorem \ref{thm-NCsingle} can be seen in Appendix \ref{sec:techdetails}. This result shows that negative controls provide the largest benefit when (i) $a^{\top}\Gamma \Gamma^{\top} b_1$ is large in magnitude, (ii) the confounding bias of the negative control given by $b_1^{\top}\check{\mathcal{G}}_j$ is large, and (iii) when $M_1^{\dagger}$ is colinear with $ \Sigma_{u|t}^{-1/2} \mu_{u|\Delta t}$. The first condition occurs when the negative control outcome and the outcome of interest have similar confounding mechanisms, meaning that the same linear combinations of the unmeasured confounders affect their outcomes. The second condition simply implies that the negative control estimand also suffers from bias due to the unmeasured confounders. The third condition occurs when the linear combinations of the unmeasured confounders that affect both the negative control treatment and the treatment of interest are similar. Therefore, ideal negative control pairs are those with large confounding biases and similar confounding mechanisms as the estimand of interest. While we focused on $J=1$ in Theorem \ref{thm-NCsingle} to build intuition, we extend this result to settings with $J>1$ in Appendix \ref{sec:techdetails}. 

\subsubsection{Identification conditions with negative controls}
\label{ssec:NCidentification}

While the focus of our paper is generally on partial identification, we show here that under certain scenarios, negative controls can point identify the causal effect of interest. This further shows the potential benefit of negative controls and provides connections between our work and identification scenarios in the literature on double negative controls \citep{miao2018confounding, shi2020multiply, shi2020selective, hu2023bayesian}. 
\begin{theorem} \label{thm-NCidentification}
Suppose we have $J$ negative control outcomes and define $b = [b_1, b_2, \dots, b_J]$ to be a $q \times J$ matrix that has each individual $b_j$ as its columns. Let $b^*$ be a $q \times J^*$ matrix comprising of a subset of the columns in $b$ where $1 \leq J^* \leq J$, and define $\mathcal{J}^*$ to be the set of indices for the columns of $b$ that make up $b^*$. Additionally, let $\mathcal{C}(A)$ represent the column space of a matrix $A$. Under the assumptions and conditions of Theorem \ref{thm-NCsingle}, the causal effect is identifiable under any of the following scenarios. 
\begin{enumerate}
    \item $a^{\top} \Gamma \in \mathcal{C}(\Gamma^{\top} b^*)$ and $b_j^{\top} \Gamma \in \mathcal{C}(M_j)$ for all $j \in \mathcal{J}^*$
    \item There exists a $j$ such that $a^{\top} \Gamma$ is colinear with $b_j^{\top} \Gamma$ and $\Sigma_{u|t}^{-1/2} \mu_{u|\Delta t} \in \mathcal{C}(M_j)$.
    \item $\Sigma_{u|t}^{-1/2} \mu_{u|\Delta t} \in \mathcal{C}(\Gamma^{\top} b^*)$ and $b_j^{\top} \Gamma \in \mathcal{C}(M_j)$ for all $j \in \mathcal{J}^*$
\end{enumerate}
\end{theorem}
A proof of this result can be found in Appendix \ref{sec:techdetails}. While these conditions may seem abstract, they do hold in some plausible scenarios. For example, scenario 1 holds automatically when we have $m$ negative control exposures, i.e. negative control pairs with the same outcome as the estimand of interest, but different exposure contrasts. 
Identification can also be obtained if additional assumptions are made, such as the existence of a so-called confounding bridge function \citep{miao2018confounding}, though we point readers to the literature on double negative controls for a discussion of such assumptions. Lastly, it is important to emphasize that within our framework, negative controls are still useful even if these conditions do not hold, as they can reduce the widths of the partial identification region substantially compared with the widths without negative controls. This is related to recent work on partial identification in the proximal causal inference literature where partial identification can be obtained when assumptions required for point identification do not hold \citep{ghassami2023partial}.

\section{Implementation details}
\label{sec:ImplementationDetails}

In this section we summarize the methodology presented in the previous sections to provide practitioners with guidance on how to implement the aforementioned approaches under a general format that works for any of the aforementioned assumptions or constraints placed on causal effects. Before we present the algorithm, it is important to first emphasize that before implementing this approach and the steps below, one should first reason through all of the necessary assumptions required for the approach to be valid, particularly the factor confounding assumption. If one believes that their treatments (or outcomes) arise in very distinct and independent ways, and therefore the shared confounding aspect of the factor confounding assumption does not hold, then they should not proceed and should consider other approaches to sensitivity analysis to the unmeasured confounding assumption. If this assumption is deemed plausible, then an analyst can follow the steps outlined in Algorithm \ref{alg:General}. To make the approach as general as possible, we can define a function of the rotation matrix, denoted by $\text{con}(R)$, which returns a TRUE/FALSE value depending on whether the constraints imposed are true or not. These constraints can contain negative control conditions, effect size constraints, partial R-squared constraints, or any other possible assumption, and can contain multiple constraints simultaneously. Note that we can write the function in this way because all of the aforementioned constraints can be written as functions of $\Gamma$ and $B$, and therefore $R$. Additionally, if no constraints are imposed and only factor confounding is assumed, then this function always returns TRUE, and the algorithm below will return the bounds in Theorem \ref{thm:bias_identifiability}.

\begin{algorithm}[H] \label{alg:General}
\SetAlgoLined
- Use any regression approach to estimate $E(T | X)$ and $E(Y | T,X)$ \\
- Select $m_1$ and $m_2$ using BIC or one of the approaches considered in Appendix C \\
- Run factor analysis with $m_1$ factors on the residuals $Y - E(Y | T,X)$ \\
\quad - Extract estimated $\Gamma_1$ and construct $\tilde \Gamma = [\Gamma_1 ~ \mathbf{0}_{q \times m_2}] $\\
- Run factor analysis with $m_2$ factors on the residuals $T - E(T | X)$ \\
\quad - Extract estimated $B_1$ and construct $B = [\mathbf{0}_{k \times m_1} ~ B_1]$ \\

- Initialize $R_{set} = \emptyset$ \\
\For{$h$ in $1, \dots, H$}{
    - Sample a rotation matrix $R^{(h)}$ from a uniform distribution over the Stiefel manifold \\
    - Check if constraints given by $\text{con}(R^{(h)})$ are true \\
    - \If{$\text{con}(R^{(h)}) = \text{TRUE}$}
    {- Add $R^{(h)}$ to the set $R_{set}$.}
    
}
- Define the partial identification region as the set of effects implied by the rotation matrices in $R_{set}$

\caption{Algorithm for finding partial identification regions}
\end{algorithm}

Note that we used sampling to find the partial identification region, though one can alternatively use optimization techniques, such as those in Section \ref{sec-NC-numerical} when appropriate. One possibility when implementing this algorithm is that the set $R_{set}$ remains empty even after sampling rotation matrices. This likely implies that one or more of the contraints are not true, factor confounding does not hold, or there exists model misspecification. Lastly, this approach only details how to find point estimates of all parameters and a single partial identification region for the causal effects of interest, but does not detail inference. One can either use the nonparametric bootstrap for inference and apply Algorithm \ref{alg:General} to each bootstrap sample, or if Bayesian inference on parameters is used, then this can be implemented within each posterior sample. Either way, a $(1 - \alpha)100\%$ partial identification region can be obtained as the smallest region that contains at least $(1 - \alpha)100\%$ of the individual partial identification regions from each bootstrap or posterior sample. We provide more details on this process in Appendix \ref{sec:AppendixEstimationFactorModel}. 

\section{Simulation study}
\label{sec:Simulation}

Here, we present a brief simulation study to highlight the benefits of characterizing the joint partial identification region for multiple estimands. For simplicity, we focus on a single data set with sample size $n=10^6$ to effectively eliminate sampling variability and focus on the uncertainty caused by the potential presence of unmeasured confounders. For brevity, we leave explicit details of the data generation process to Appendix \ref{sec:AppendixSimDetails}. To summarize, we consider a problem with $k=10$ exposures and $q=6$ outcomes, with $m=3$ unmeasured confounders. The true effect sizes for all estimands are between $-1$ and 1, and there are a number of treatments with no effect on specific outcomes, leading to multiple potential negative controls. We use linear models throughout, which contains the true data generating mechanism, though similar results hold for more complex estimators of $g(\cdot)$. Our inferential goal is to estimate the effect of a particular exposure on a single outcome. We scaled the confounding and effect sizes so that the estimate obtained assuming no unmeasured confounding is 0, while the true value of the estimand is 1. To illustrate the different assumptions or constraints that one can place, we derive the partial identification region for this estimand under a variety of different assumptions, summarized below. 
\begin{enumerate}
    \item Bounds in equation \eqref{interval} that rely only on the factor confounding structure (Factor confounding) 
    \item Incorporating a single negative control outcome (Negative control outcome)
    \item Incorporating a negative control for which the treatment and outcome for the negative control are both different from the estimand of interest  (Negative control pair)
    \item Incorporating both negative control variables simultaneously (Both negative controls)
    \item Effect size constraints as in Section \ref{ssec:BoundsTreatmentEffects} with $z_l = -1.2$ and $z_u = 1.2$. Constraints are simultaneously placed on the effect of the exposure of interest on the five outcomes that are not of interest. (Effect size $< 1.2$)
    \item The same effect size constraints, but with $z_l = -2$ and $z_u = 2$ (Effect size $< 2$)
    \item Partial R-squared constraints as in Section \ref{ssec:PartialRsquaredBounds} where we constrain $R^2_{d^{\top}T \sim U_{a,d}} < 0.2$ for the outcome of interest and all corresponding exposure contrasts. 
\end{enumerate}

\begin{figure}[h]
    \centering
    \includegraphics[width=0.95\linewidth]{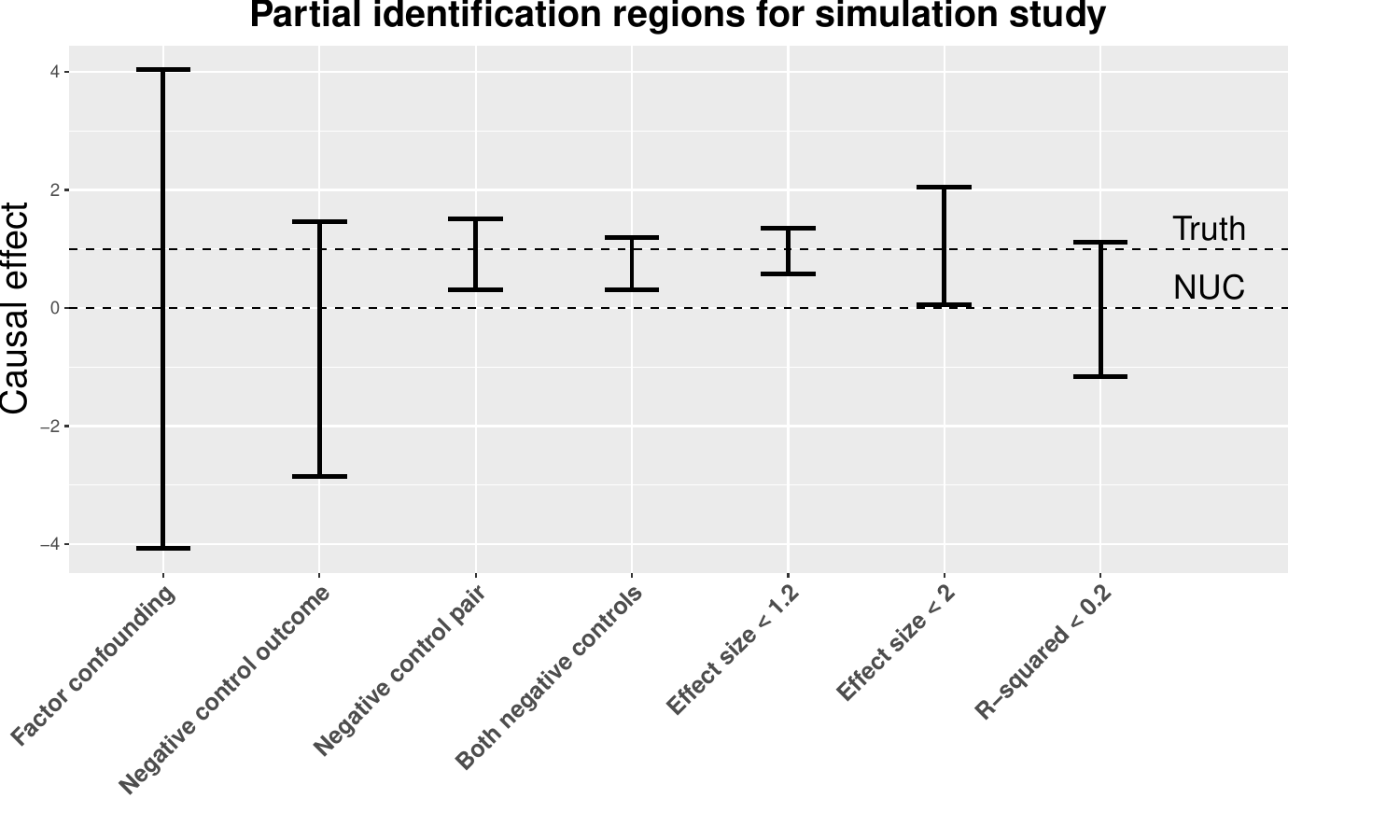}
    \caption{Partial identification regions under different constraints in simulated data set. The true value is represented by the dotted line at 1, while the estimate obtained assuming no unmeasured confounding (NUC) is represented by the dotted line at 0.}
    \label{Fig:SimulationStudyPartialID}
\end{figure}

The results can be found in Figure \ref{Fig:SimulationStudyPartialID}, where we see that all partial identification regions contain the true value of the estimand, though different constraints lead to very different partial identification regions. In terms of negative controls, we see in this case that the negative control pair, which looks at a completely distinct treatment and outcome from the estimand of interest, is still very informative for the estimand of interest, as they have similar confounding mechanisms. Having two negative control exposures is also very informative, and unsurprisingly leads to the smallest intervals among the different negative control assumptions. The effect size constraints are extremely powerful in this case as well, even for the more relaxed constraint using 2 as a cutoff value, as the region is very small and contains the true value.  Overall, these results show that different constraints or assumptions can be more or less informative about the estimand of interest depending on the situation, and that using no constraints (factor confounding only) can lead to very wide, uninformative intervals. Reasonable additional constraints result in far more informative intervals, some of which are correctly able to rule out a causal effect of zero, despite zero being the estimate obtained assuming no unmeasured confounding. This highlights that the approaches developed here need not be centered around the estimate obtained with no unmeasured confounding, and that the partial identification regions are able to provide tight bounds around the true value, even under a large amount of confounding bias. 

\section{Estimating the health effects of components of PM$_{2.5}$}
\label{sec:Application}

Now we return to the motivating example and data analysis, which was introduced in Section \ref{sec:DataDescription}. We standardize all exposures and outcomes to have mean zero and standard deviation one prior to running the analysis so that the effect estimates are all on the same scale across outcomes and exposure combinations. For example, an estimate of 0.1 from any exposure shift corresponds to a 0.1 standard deviation change in the outcome. One distinction between our data and the general framework presented throughout the manuscript is that we have repeated observations over time. Specifically, we observe $n=30,448$ zip codes across a seventeen year period from 2000 to 2016. We treat each year and zip code combination as a distinct data point by stacking data from each year into one final data set, though all uncertainty quantification is done using a cluster-level bootstrap at the zip code level to ensure that dependence in the observations is accounted for. Our main estimand is therefore
$$E[a^\top Y_{is}(t_1) - a^\top Y_{is}(t_2)],$$
where $Y_{is}$ represents the outcome for zip code $i$ in year $s$, and the expectation is over both time and units. Note that there is a possibility to combine the partial identification framework seen here with panel data type approaches that explicitly leverage the repeated observations over time to account for confounding \citep[see e.g][]{wu2026latentfactorpanelapproach}. Such an approach is beyond the scope of this paper and presents an interesting avenue for future research. In order to evaluate whether the causal effect varies over time, we also consider time-specific estimates where the averaging is only done over units at a specific point in time. We consider a range of values for the vector $a$ that correspond to different public health outcomes, and we let $t_1$ and $t_2$ be values of the exposure vector that correspond to single-component shifts in the air pollution mixture. All values in $t_1$ and $t_2$ are set to the median exposure value except for the exposure being studied, for which the 1st and 3rd quartiles are used. We estimate $g(\cdot)$ and therefore the effects of the exposures (and confounders) on the outcome using a gradient boosting regression model to ensure that nonlinearities and interactions between exposures are accounted for, and use the aforementioned approaches to assess the robustness of these estimates to unmeasured confounding bias. In what follows, we present partial identification regions for the causal effects of the air pollution mixture under assumptions about effect sizes (Section \ref{ssec:MedicareEffectSize}), or assumptions about the strength of confounding (Section \ref{ssec:MedicareConfoundingStrength}). Throughout, we show results when $m_1 = 3$ and $m_2 = 3$ as selected by BIC. Additional results, such as exploratory analyses, estimated factor loadings, analyses not assuming factor confounding for the outcome, and causal effects for estimands not considered here can be found in Appendix \ref{sec:AppendixExtraMedicare}.

\subsection{Plausibility of key assumptions}
\label{ssec:MedicareAssumptions}

Before showing partial identification regions for the causal effects of interest, it is important to first examine the plausibility of key assumptions that our approach is reliant upon. The core assumption for our approach rests on factor confounding (Assumption \ref{ass-iden4}), which effectively states that the unmeasured confounders are associated with multiple exposures and multiple outcomes. We believe that it is plausible this assumption holds, at least approximately, within our application for a number of reasons. All outcomes considered in our study correspond to hospitalization rates for different diseases, and it is reasonable that certain characteristics, which are not measured in our observed covariates, are broadly associated with poor health outcomes. For instance, these unmeasured characteristics likely reflect features like health-seeking behavior or proximity to and/or quality of health care access. Similarly, exposures tend to be elevated in areas with various pollutant emitting sources such as roadways, power plants, and factories. Certain communities are more likely to be near these sources of pollution due to underlying structural inequalities. Further, multiple exposures can be elevated due to similar climatic or atmospheric processes. All of these point to the plausibility of the shared confounding mechanism within the exposures considered. Our results also rely on a Gaussian probablistic principal components model for the exposures (Equations \ref{eqn:u} and \ref{eqn:t}) and homoskedasticity of the latent confounders (Assumption \ref{asm:gaussian_latent}).  While this appears to be a strong assumption, related work on accounting for unmeasured confounding with Gaussian latent variable models show that, in practice, results are quite robust to departures from normality and homoskedasticity  \citep{wu2026latentfactorpanelapproach}. As such, even when these assumptions do not hold exactly, our approach can be effective at capturing the first order effects of confounding bias.

Finally, we briefly comment on the plausiblity of additional standard causal assumptions like positivity and SUTVA. While latent positivity, which conditions on both $X$ and $U$ can not be tested empirically, we can assess positivity conditional on $X$. While there is not a well-accepted approach to assessing positivity for a multivariate, continuous exposure vector, analyses using the same data source seen here have shown that positivity holds at least approximately \citep{shin2023spatial}. The SUTVA assumption could be violated if the exposure for some zip codes affect outcomes in other zip codes, which could occur due to mobility of individuals across zip codes. A recent analysis of air pollution studies using cell phone level mobility information studied this prospect and showed that effect estimates are not largely impacted by ignoring such mobility, and if anything, accounting for mobility would lead to larger point estimates of causal effects and estimates that are more robust to unmeasured confounding \citep{shin2023spatial}. For this reason, we believe it is reasonable to assume SUTVA in this setting.

\subsection{Results using effect size constraints}
\label{ssec:MedicareEffectSize}

We now explore the partial identification regions for the effect of both organic carbon and nitrates on the outcomes of interest when using either negative control or other effect size constraints. We focus on these two exposures in this section for simplicity, and because these two exposures illustrate two distinct scenarios where either the partial identification regions are informative about the causal effect, or they are not informative. The results for other exposures are included in Appendix \ref{sec:AppendixExtraMedicare}. The partial identification regions for the other exposures generally contain zero and are wide for elemental carbon, sulfates, and ammonium due to the large partial R-squared values seen in Table 1 in Appendix \ref{sec:AppendixExtraMedicare}. We incorporate two different constraints when calculating the partial identification regions for each exposure of interest: 1) using prior COPD as a negative control outcome, and 2) placing an effect size constraint on the effect of that exposure on asthma hospitalizations with $z_l = -0.03$ and $z_u = 0.1$. The lower bound stems from the implausibility of pollution having a protective effect on health, while the upper bound is based on prior literature looking at larger shifts in particulate matter than those considered here, which have estimated smaller effects than this upper bound \citep{di2017air}. Complex public health outcomes such as hospitalization rates are driven by a host of factors (smoking and obesity, among others), and particulate matter can only explain a small part of the variability in these outcomes. Given that our outcomes are standardized, it is therefore unlikely to have an effect size greater than 0.1 as this would imply that pollution explains a moderately large portion of the variability in hospitalization rates. 

\begin{figure}[h]
    \centering
    \includegraphics[width=0.9\linewidth]{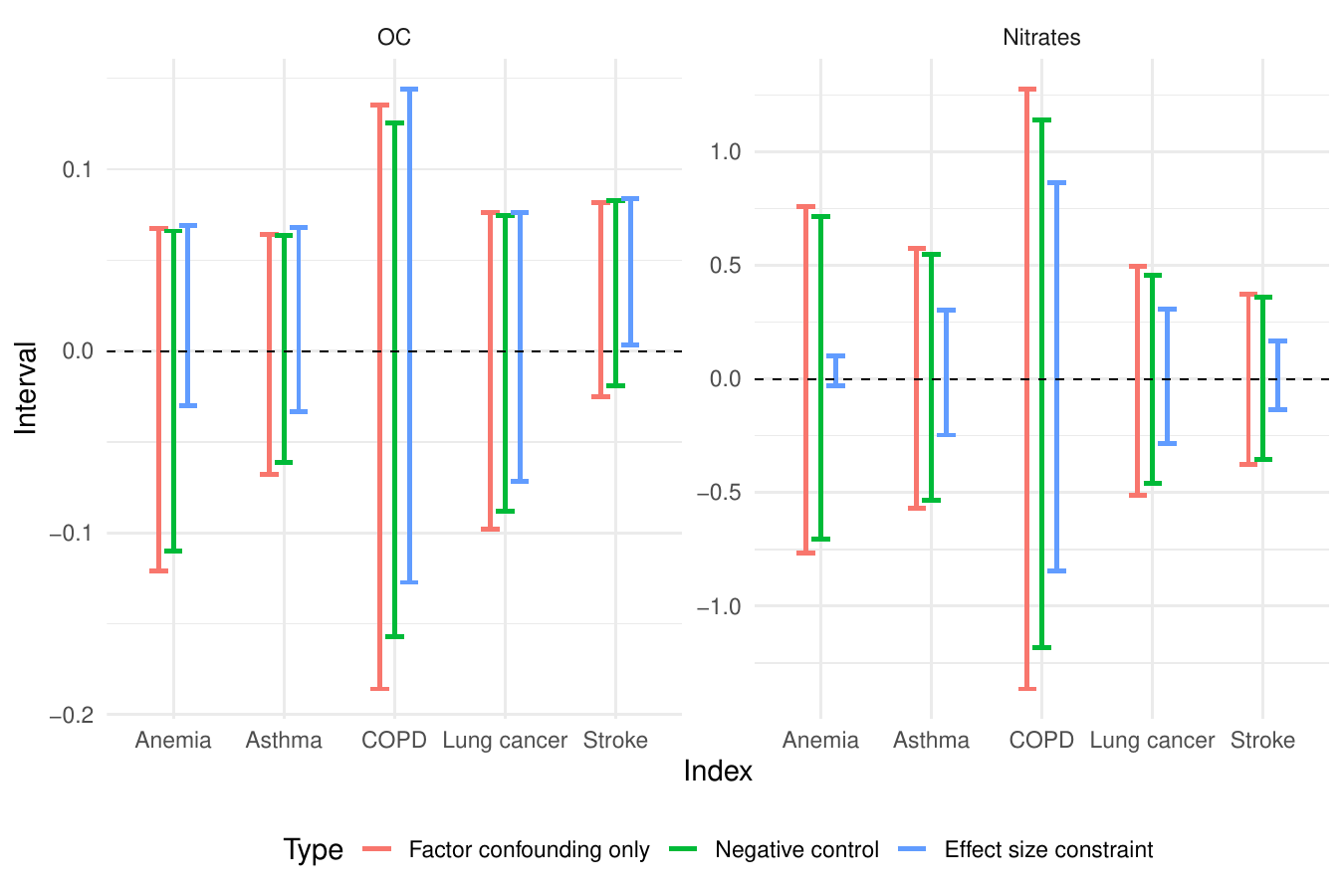}
    \caption{Partial identification regions for the effect of organic carbon and nitrates on all health outcomes of interest. }
    \label{Fig:MedicareOC_Nitrates}
\end{figure}

\begin{figure}[h]
    \centering
    \includegraphics[width=0.9\linewidth]{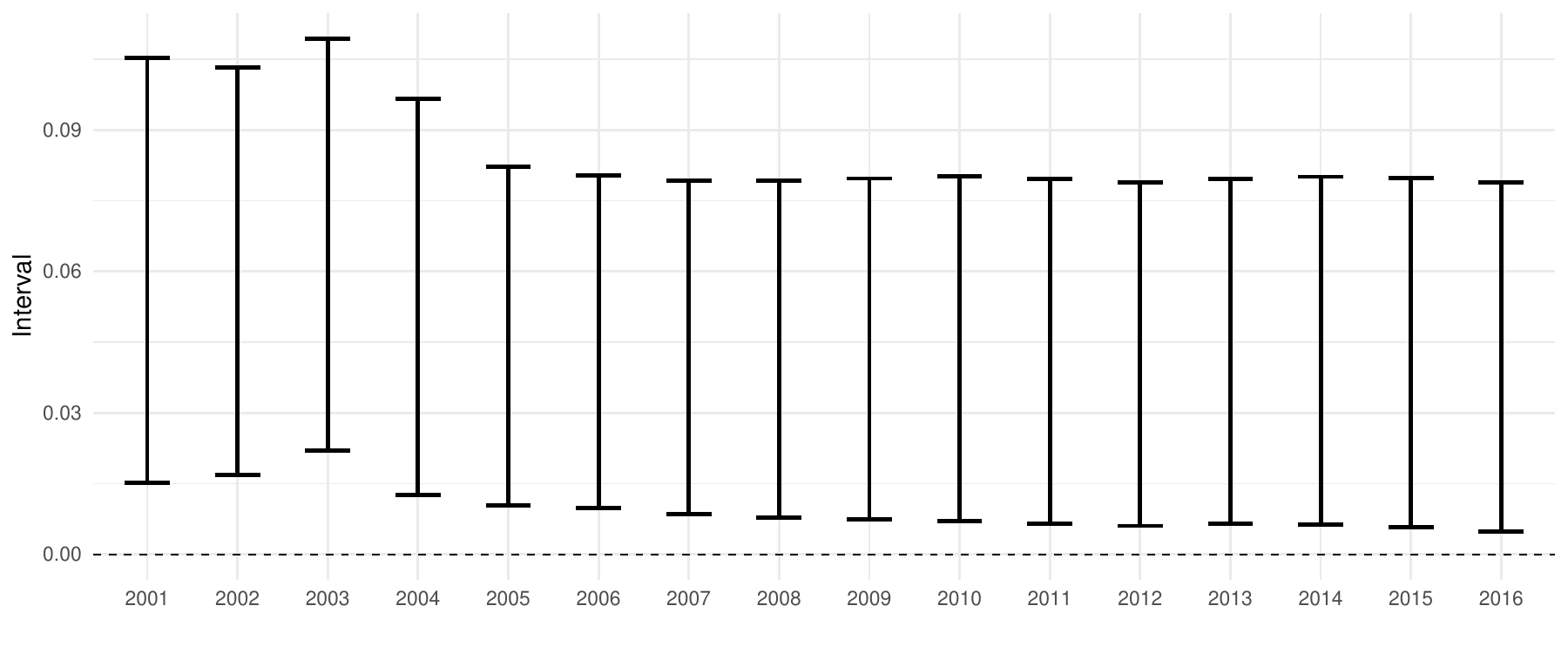}
    \caption{Estimated causal effects and partial identification regions for the causal effect of organic carbon on stroke hospitalization rates over time.}
    \label{Fig:MedicareOC_Yearly}
\end{figure}

The results for both exposures can be found in Figure \ref{Fig:MedicareOC_Nitrates}. The intervals are generally wider and less informative for nitrates, likely due to the larger value of $R^2_{d^{\top}T \sim U | X}$ for this exposure (see Appendix F). The intervals are more informative for organic carbon though for most outcomes they contain zero and do not rule out the possibility of there being no causal effect of organic carbon. The one exception is for stroke hospitalization rates, for which the partial identification regions lie entirely above 0 under the effect size constraint assumption. We can also study causal effects over time to examine the extent to which these effects have changed over the course of the study. Figure \ref{Fig:MedicareOC_Yearly} shows the effect of organic carbon on stroke hospitalization rates across the years of our study. We see that the effects do not vary greatly over time, though are generally larger and more pronounced in earlier years of the study. 

\begin{figure}[h]
    \centering
    \includegraphics[width=0.9\linewidth]{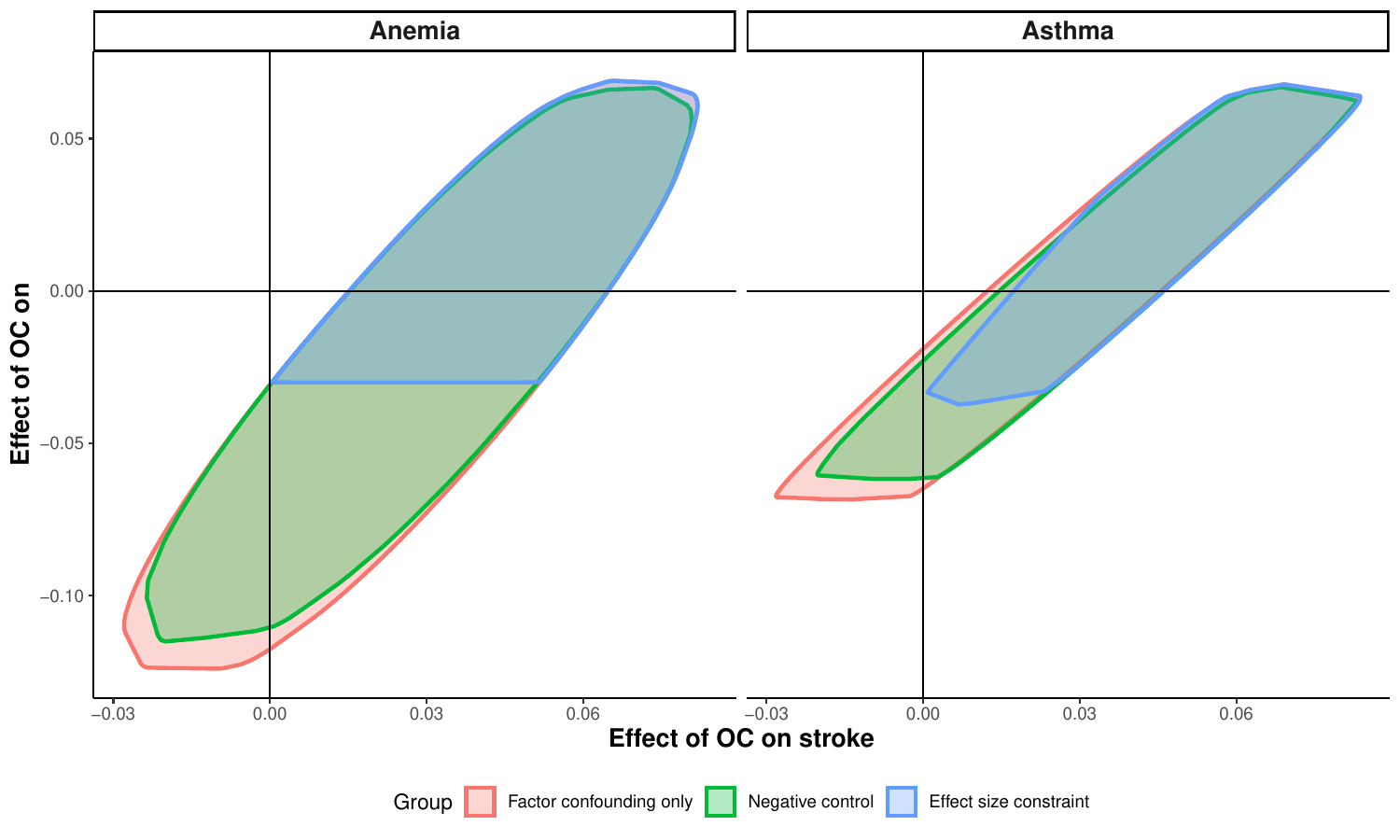}
    \caption{Bivariate partial identification regions for the effect of organic carbon on stroke and either anemia or asthma hospitalization rates. }
\label{Fig:MedicareBivariatePlots}
\end{figure}

As discussed in Section \ref{ssec:MultipleEstimands}, more refined information about the plausible values of these causal effects can be seen by examining bivariate partial identification regions. For example, in Figure \ref{Fig:MedicareBivariatePlots}, we see that the lower bound for the effect of organic carbon on stroke rates can be sharpened depending on one's belief of the effect of organic carbon on anemia or asthma hospitalizations. For example, if one assumes there is no effect of organic carbon on anemia, then the effect on stroke is clearly above zero and much larger than the lower bound seen in the univariate region for the effect of organic carbon on stroke rates. The minimum value of the effect of organic carbon on stroke rates is only obtained if there is a large, protective effect of organic carbon on anemia and asthma as well, which is not likely. Another interesting finding from Figure \ref{Fig:MedicareBivariatePlots} is understanding how the effect size constraint is able to sharpen inferences substantially. By ensuring that there is not a large protective effect of organic carbon on anemia (since $z_l = -0.03$), the partial identification regions become much more informative for the effect of organic carbon on both stroke and asthma hospitalization rates.  Overall, this analysis provides moderately strong, robust evidence of a causal effect of organic carbon on stroke hospitalization rates.

\subsection{Results using assumptions on strength of confounding}
\label{ssec:MedicareConfoundingStrength}

\begin{figure}[h]
    \centering
    \includegraphics[width=0.9\linewidth]{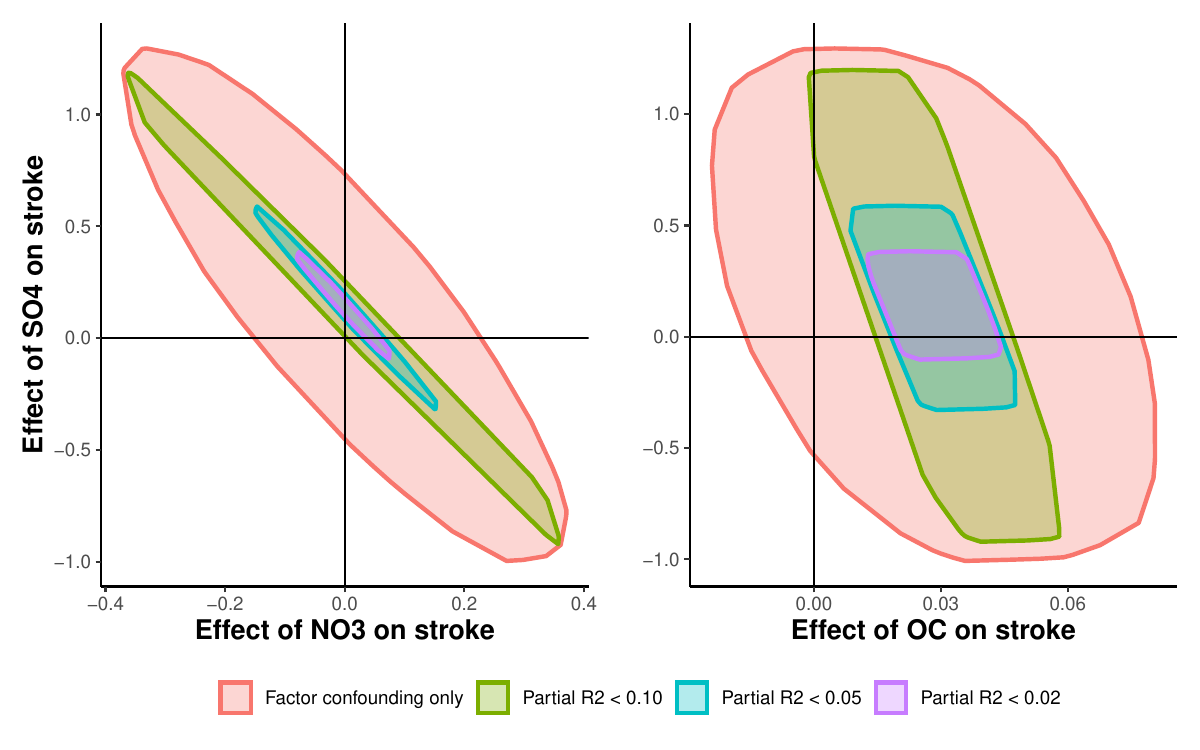}
    \caption{Partial identification regions for the effects of nitrates, sulfates, and organic carbon on stroke hospitalization rates when making assumptions about the strength of confounding.}
    \label{Fig:MedicareR2results}
\end{figure}

In this section, we explore alternative assumptions focusing on the degree of confounding as opposed to assumptions about effect sizes. We again examine the effects of organic carbon and nitrates, and we focus on their effect on stroke hospitalization rates to see if the effects seen in the previous section continue to hold under an alternative set of assumptions. We also examine the effect of sulfates on stroke hospitalization rates to explore whether these assumptions can provide more informative bounds than effect size constraints, which led to overly wide intervals for the effect of sulfates (see Appendix \ref{sec:AppendixExtraMedicare}). We use the methodology described in Section \ref{ssec:PartialRsquaredBounds} and we restrict $R^2_{d^{\top}T \sim U_{a,d}} < r_l$ for all exposures. To explore differing degrees of confounding, we examine $r_l \in \{0.02, 0.05, 0.1 \}.$ To put these magnitudes into context, if we use the formal benchmarking framework of \cite{cinelli2020making} and assume that the unmeasured variable is as strongly correlated with the exposures as the strongest observed covariate, then the largest benchmarking value of $R^2_{d^{\top}T \sim U_{a,d}}$ that we obtain for any exposure is approximately 0.05. This shows that we are exploring associations with the exposure that are weaker, equal to, or twice as strong as the strongest observed covariate. Also note that our choice of deconfounder used in Section \ref{ssec:PartialRsquaredBounds} maximizes the correlation between the unmeasured variable and the outcome that is allowed under factor confounding. The results from this analysis can be found in Figure \ref{Fig:MedicareR2results}, which shows bivariate partial identification regions for the effects of organic carbon, nitrates, and sulfates on stroke hospitalization rates. From the right panel, we can see that the effect of organic carbon is fully contained above zero when the partial R-squared values are less than 0.1. Even when the partial R-squared value is 0.1, the effect of organic carbon on stroke hospitalizations is only zero when the effect of sulfates on stroke hospitalizations is overly large. This shows that even with a very large degree of unmeasured confounding, there is evidence of an effect of organic carbon on stroke hospitalizations. The left panel shows the effects of nitrates and sulfates on stroke hospitalizations, and these two exposures generally have much larger partial identification regions. Despite this, our results here point to there being an effect of one, or both, of these exposures. The bivariate partial identification region only includes the point (0,0) when the partial R-squared is set to 0.1, and anything smaller than that leads to partial identification regions that do not contain this null value. Overall, these results show that moderate degrees of unmeasured confounding are not sufficient to fully remove the effects of particulate matter components on stroke hospitalizations. 

\section{Discussion}

In this paper, we develop a framework for identifying partial identification regions in the presence of unmeasured confounding when the data contain multiple treatments and multiple outcomes, which is particularly useful for studying the health effects of multivariate air pollutants. We illustrate how a factor confounding assumption can be leveraged to learn more about the range of plausible causal effects across multiple estimands. For one, we obtain bounds on causal effects that are identifiable from the observed data alone. These partial identification regions can be reduced in size significantly under reasonable constraints that are known a priori given subject-matter expertise, which one typically has in environmental studies. One of the constraints we consider are negative control assumptions, for which we develop theory and algorithms to produce refined partial identification regions. Alternatively, prior knowledge on plausible effect sizes or the strength of confounding can be used to provide more informative partial identification regions. Overall, our theoretical and empirical studies suggest that the proposed framework yields new insights about causal effects even in the presence of unmeasured confounding variables. This was evident in a nationwide analysis of the effects of particulate matter components on cause specific hospitalizations using Medicare claims data, which showed harmful effect of particulate matter on public health outcomes.

There are a number of potentially interesting future research directions that could expand upon this work and make partial identification in multi-treatment and multi-outcome settings more widely applicable. For one, we have focused on continuous treatments here, which facilitates the use of the factor models used throughout, but extending these ideas to categorical treatments and generalized linear models would be an important extension of this work. Related, our current framework relies on the factor confounding assumption, and it would be useful to study the robustness of our findings to violations of this assumption. Lastly, we studied the refinement of the partial identification region in the presence of negative controls or prior constraints on effect sizes, but other a priori assumptions, such as sparsity of treatment effects, could be incorporated as well. Extending our work to these other situations would be a natural extension that could broaden the scope of this work.

\section*{Acknowledgements}

Research described in this article was conducted under contract to the Health Effects Institute (HEI), an organization jointly funded by the United States Environmental Protection Agency (EPA) (Assistance Award No. CR-83590201) and certain motor vehicle and engine manufacturers. The contents of this article do not necessarily reflect the views of HEI, or its sponsors, nor do they necessarily reflect the views and policies of the EPA or motor vehicle and engine manufacturers. The computations in this paper were run on the FASRC Cannon/FASSE cluster supported by the FAS Division of Science Research Computing Group at Harvard University.

\bibliographystyle{apalike}
\bibliography{bibfile.bib} 

\appendix

\newpage
\section{Technical Details} \label{sec:techdetails}
\subsection{Proofs of Proposition 1, and Theorems 1, 2, 3, and 4} 

\begin{proof}[Proof of Theorem 1]\label{pf-thm-biasbound}
The omitted confounding bias is bounded as follows:
\begin{align*}
    \text{Bias}_{a,t_1,t_2}^2 
    &=  \left( a^{\top}\Gamma \Sigma_{u|t}^{-1/2} \mu_{u|\Delta t} \right)^2 \\
    &= \|a^{\top}\Gamma\|_2^2 \| \Sigma_{u|t}^{-1/2} \mu_{u|\Delta t} \|_2^2 \cos^2(\theta_{a,t_1,t_2}) \\
    &\leq \|a^{\top}\Gamma\|_2^2 \| \Sigma_{u|t}^{-1/2} \mu_{u|\Delta t} \|_2^2,
\end{align*}
where $\theta_{a,t_1,t_2}$ represents the angle between the vectors $a^{\top}\Gamma$ and $\Sigma_{u|t}^{-1/2} \mu_{u|\Delta t}$.
This completes the proof, and the bound can be reached when $a^{\top}\Gamma$ is colinear with $\Sigma_{u|t}^{-1/2} \mu_{u|\Delta t}$.
\end{proof}

\begin{proof}[Proof of Theorem 2]\label{pf-prop-iden}
We first show the identification of $\|a^{\top}\Gamma\|_2$. Under the first condition of Assumption 4, $\Gamma\Gamma^{\top}$ and $\Sigma_{y|t,u}$ is uniquely determined from
\begin{align*}
    \text{Cov}[Y-g(T)|T=t]
    &= \Gamma\Sigma_{u|t}^{-1/2}\text{Cov}(U|T=t)(\Gamma\Sigma_{u|t}^{-1/2})^{\top} + \text{Cov}(\epsilon_Y) \\
    &= \Gamma\Sigma_{u|t}^{-1/2}\Sigma_{u|t}\Sigma_{u|t}^{-1/2}\Gamma^{\top} + \Sigma_{y|t,u} \\
    &= \Gamma\Gamma^{\top} + \Sigma_{y|t,u},
\end{align*}
by applying Lemma 5.1 and Theorem 5.1 of \citet{anderson1956statistical}, $\Gamma$ is identified \textit{up to rotations} from the right under the factor confounding assumption. 
In other words, $\Gamma$ is identified up to multiplication on the right by an orthogonal matrix, and thus any admissible value for $\Gamma$ can be written as $\widetilde{\Gamma}=\Gamma R$ with an arbitrary $m \times m$ orthogonal matrix $R$. Then $\|a^{\top}\Gamma\|_2$, which is rotation-invariant, is identified as follows.
\begin{align*}
    \|a^{\top}\widetilde{\Gamma}\|_2^2
    &= \|a^{\top}\Gamma R\|_2^2 \\
    &= a^{\top}\Gamma R(\Gamma R)^{\top}a \\
    &= a^{\top}\Gamma\Gamma^{\top}a \\
    &= \|a^{\top}\Gamma\|_2^2,
\end{align*}
since $RR^{\top}=I$.

Next, we show the identification of $\| \Sigma_{u|t}^{-1/2} \mu_{u|\Delta t} \|_2$, by applying the same idea to the treatment model. Under condition 2 of Assumption 4, $BB^{\top}$ and $\Sigma_{t|u}$ are uniquely determined from $\text{Cov}(T)=BB^{\top}+\text{Cov}(\epsilon_t)=BB^{\top}+\Sigma_{t|u}$, and any admissible value for $B$ can be written as $\widetilde{B}=B R$ where $R$ is again an arbitrary $m \times m$ orthogonal matrix. Then, the conditional distribution of $U$ given $T$ with $\widetilde{B}$ is
\begin{equation*}
    f_{\widetilde{B}}(u|T=t) \sim N_m(\widetilde{\mu}_{u|t}, \widetilde{\Sigma}_{u|t}),
\end{equation*}
with
\begin{align*}
    \widetilde{\mu}_{u|t}
    &= \widetilde{B}^{\top}(\widetilde{B}\widetilde{B}^{\top}+\Sigma_{t|u})^{-1}t \\
    &= R^{\top}B^{\top}(BB^{\top}+\Sigma_{t|u})^{-1}t \\
    &= R^{\top}\mu_{u|t}, \\
    \widetilde{\Sigma}_{u|t}
    &= I_m - \widetilde{B}^{\top}(\widetilde{B}\widetilde{B}^{\top}+\Sigma_{t|u})^{-1}\widetilde{B} \\
    &= I_m - R^{\top}B^{\top}(BB^{\top}+\Sigma_{t|u})^{-1}BR \\
    &= R^{\top}\Sigma_{u|t}R.
\end{align*}
Then, we can show that
\begin{align*}
    \| \widetilde{\Sigma}_{u|t}^{-1/2} \widetilde{\mu}_{u|\Delta t} \|^2_2
    &= \| (R^{\top}\Sigma_{u|t}R)^{-1/2} (R^{\top}\mu_{u|\Delta t}) \|^2_2 \\
    &= \text{tr}\left\{(R^{\top}\mu_{u|\Delta t})^{\top} (R^{\top}\Sigma_{u|t}R)^{-1} (R^{\top}\mu_{u|\Delta t})\right\} \\
    &= \text{tr}\left( \mu_{u|\Delta t}^{\top} \Sigma_{u|t}^{-1} \mu_{u|\Delta t} \right) \\
    &= \| \Sigma_{u|t}^{-1/2} \mu_{u|\Delta t} \|^2_2,
\end{align*}
where $RR^{\top}=I$, which implies that $\| \Sigma_{u|t}^{-1/2} \mu_{u|\Delta t} \|_2$ is identified. 
Combining the identification results, the bias bound is identifiable under the conditions of Assumption 4.

Lastly, we show that even though both $B$ and $\Gamma$ are only identifiable up to rotation, an equivalent representation of the bias allows us to treat $\widetilde{B}$ as the true value $B$ and only focus on identifying the rotation matrix associated with $\Gamma$. Before doing this, let us first define the eigendecomposition $\Sigma_{u|t} = Q A Q^{-1}$, which implies that $\Sigma_{u|t}^{-1/2} = Q A^{-1/2} Q^{-1}$. We also define two separate rotation matrices for this calculation. We let $R_1$ be the rotation for $\Gamma$ such that $\widetilde{\Gamma} = \Gamma R_1$, and define $R_2$ so that $\widetilde{B} = B R_2$. Using the relationships described above, it is easy to show that $\widetilde{\Sigma}_{u|t}^{-1/2} = R_2^{\top} Q A^{-1/2} Q^{-1} R_2 = R_2^{\top} \Sigma_{u|t}^{-1/2} R_2$. Putting this all together, we can write the bias as
\begin{align*}
    a^{\top} \Gamma \Sigma_{u|t}^{-1/2} \mu_{u|t} &= a^{\top} \widetilde{\Gamma} R_1 \Sigma_{u|t}^{-1/2} \mu_{u|t} \\
    &= a^{\top} \widetilde{\Gamma} R_1 R_2 \widetilde{\Sigma}_{u|t}^{-1/2} R_2^{\top} \mu_{u|t} \\
    &= a^{\top} \widetilde{\Gamma} R_1 R_2 \widetilde{\Sigma}_{u|t}^{-1/2} R_2^{\top} R_2 \widetilde{\mu}_{u|t} \\
    &= a^{\top} \widetilde{\Gamma} R_1 R_2 \widetilde{\Sigma}_{u|t}^{-1/2} \widetilde{\mu}_{u|t} \\
    &= a^{\top} \widetilde{\Gamma} R^* \widetilde{\Sigma}_{u|t}^{-1/2} \widetilde{\mu}_{u|t}
\end{align*}
where $R^* = R_1 R_2$ is an orthogonal matrix, since the product of orthogonal matrices is also orthogonal. This shows that we can treat the $\widetilde{B}$, $\widetilde{\Sigma}_{u|t}^{-1/2}$, and $\widetilde{\mu}_{u|t}$ parameters as the true values, and the only parameter off by rotation is $\widetilde{\Gamma}$. 

\end{proof}

\begin{proof}[Proof of Proposition 1 and Theorem 3]\label{pf-thm-NCsingle}
Here we focus on Theorem 3, but the result in Proposition 1 can be found as a special case of the result in Theorem 3 and therefore we do not provide an explicit proof. We assume that we only have a single negative control contrast. This provides us information about the plausible values of $b_j^{\top}\Gamma$.
To ensure that the negative control assumptions are compatible, the solution for (16) exists as
\begin{equation} \label{bGamma0}
    b_j^{\top}\Gamma = b_j^{\top}\check{\mathcal{G}}_j M_{j}^{\dagger} + w_{b,j} (I - M_{j} M_{j}^{\dagger}),
\end{equation}
for some arbitrary $m$-dimensional row vector $w_{b,j}$, if and only if
\begin{equation*}
    b_j^{\top}\check{\mathcal{G}}_j M_{j}^{\dagger} M_{j} = b_j^{\top}\check{\mathcal{G}}_j
\end{equation*}
holds \citep{penrose1955generalized}, where $M_{j}^{\dagger}$ denotes a generalized inverse of $M_{j}$, $M_{j} M_{j}^{\dagger}$ is the projection matrix onto the column space of $M_{j}$, and $M_{j}^{\dagger} M_{j}$ is the projection matrix onto the row space of $M_{j}$. 
Note here that if $c_j = m$, we have that $M_{j}^{\dagger}$ is a standard inverse instead of a generalized inverse, and we are able to identify $b_j^{\top}\Gamma$.

We also have a different piece of information about the relationship between $a^{\top}\Gamma$ and $b_j^{\top}\Gamma$, which we have from our factor model assumptions. This is given by
\begin{equation} \label{kab}
    b_j^{\top}\Gamma \Gamma^{\top} a = a^{\top}\Gamma \Gamma^{\top} b_j = k_{ab}
\end{equation}
for some constant $k_{ab}$ that is identified from our factor model assumptions. We can combine these two pieces of information to see that 
\begin{align} \label{aGamma0}
    a^{\top} \Gamma &= k_{ab} ({D}^{\top})^{\dagger} + w_a \{I - ({D}^{\top})({D}^{\top})^{\dagger}\},
\end{align}
where
\begin{equation*}
    D := b_j^{\top}\Gamma = b_j^{\top}\check{\mathcal{G}}_j M_{j}^{\dagger} + w_{b,j} (I - M_{j} M_{j}^{\dagger})
\end{equation*}
is an $m$-dimensional row vector.
Since $D$ is a vector, we here use the fact that
\begin{equation*}
    D^{\dagger} = \frac{D^{\top}}{\| D \|_2^2} = \frac{D^{\top}}{\| b_j^{\top} \Gamma \|_2^2} = \frac{D^{\top}}{k_{bb}}~~~\text{and thus}~~~(D^{\dagger})^{\top} = (D^{\top})^{\dagger} = \frac{D}{k_{bb}},
\end{equation*}
where $k_{bb} := b_j^{\top}\Gamma \Gamma^{\top} b_j$ is an identifiable constant.

Combining these ideas together, for treatment contrast $\Delta t = t_1-t_2$, we can see that the omitted variable bias of $\text{PATE}_{a,t_1,t_2}$ is
\begin{align*}
   \text{Bias}_{a,t_1,t_2} &= a^{\top}\Gamma \Sigma_{u|t}^{-1/2} \mu_{u|\Delta t} \nonumber \\
    &= k_{ab} ({D}^{\top})^{\dagger} \Sigma_{u|t}^{-1/2} \mu_{u|\Delta t} + w_a \{I - ({D}^{\top})({D}^{\top})^{\dagger}\} \Sigma_{u|t}^{-1/2} \mu_{u|\Delta t} \nonumber  \\
    &= \frac{k_{ab}}{k_{bb}} b_j^{\top}\Gamma \Sigma_{u|t}^{-1/2} \mu_{u|\Delta t} + w_a \{I - ({D}^{\top})({D}^{\top})^{\dagger}\} \Sigma_{u|t}^{-1/2} \mu_{u|\Delta t} \nonumber \\
    &= \underbrace{\frac{k_{ab}}{k_{bb}} b_j^{\top}\check{\mathcal{G}}_j M_{j}^{\dagger} \Sigma_{u|t}^{-1/2} \mu_{u|\Delta t}}_{(I) \text{ bias correction}} + \underbrace{\frac{k_{ab}}{k_{bb}} w_{b,j} (I - M_{j} M_{j}^{\dagger}) \Sigma_{u|t}^{-1/2} \mu_{u|\Delta t}}_{(II)} \\
    &~~~+ \underbrace{w_a \{I - ({D}^{\top})({D}^{\top})^{\dagger}\} \Sigma_{u|t}^{-1/2} \mu_{u|\Delta t}}_{(III)}
\end{align*}

The first term (I) is identifiable and is the \textit{bias correction term} for the center of our new interval. 
Now all that is left to do is to bound $(II)$ and $(III)$, which include the free vectors $w_a$ and $w_{b,j}$ and are therefore unidentifiable. 

We can first bound $(II)$. We can see that the $l_2$ norm of this vector is given by
\begin{align*}
    &~~~\left\| \frac{k_{ab}}{k_{bb}} w_{b,j} (I - M_{j} M_{j}^{\dagger}) \Sigma_{u|t}^{-1/2} \mu_{u|\Delta t} \right\|_2^2 \\
    &= \left\| \frac{k_{ab}}{k_{bb}} w_{b,j} (I - M_{j} M_{j}^{\dagger})^2 \Sigma_{u|t}^{-1/2} \mu_{u|\Delta t} \right\|_2^2 \\
    &\leq \left\| \frac{k_{ab}}{k_{bb}} w_{b,j} (I - M_{j} M_{j}^{\dagger}) \right\|_2^2  \left\| (I - M_{j} M_{j}^{\dagger})\Sigma_{u|t}^{-1/2} \mu_{u|\Delta t} \right\|_2^2,
\end{align*}
where the first equality holds because $(I - M_{j} M_{j}^{\dagger})$ is an idempotent matrix. Taking the $l_2$ norm of both sides of \eqref{bGamma0} and re-arranging terms gives us that
\begin{equation*}
    \| w_{b,j} (I - M_{j} M_{j}^{\dagger}) \|_2^2 = \| b_j^{\top} \Gamma \|_2^2 - \| b_j^{\top}\check{\mathcal{G}}_j M_{j}^{\dagger} \|_2^2.
\end{equation*}
Hence,
\begin{align*}
    |(II)|
    &\leq \left\| \frac{k_{ab}}{k_{bb}} w_{b,j} (I - M_{j} M_{j}^{\dagger}) \right\|_2  \left\| (I - M_{j} M_{j}^{\dagger})\Sigma_{u|t}^{-1/2} \mu_{u|\Delta t} \right\|_2 \\
    &= \Big| \frac{k_{ab}}{k_{bb}} \Big| \sqrt{k_{bb} - \| b_j^{\top}\check{\mathcal{G}}_j M_{j}^{\dagger} \|_2^2} \left\| (I - M_{j} M_{j}^{\dagger})\Sigma_{u|t}^{-1/2} \mu_{u|\Delta t} \right\|_2,
\end{align*}
where we use the notation $k_{bb} = \| b_j^{\top} \Gamma \|_2^2$.

In a similar way, we can get an upper bound of $(III)$. We here use the fact that
\begin{align*}
    &~~~\| w_a \{I - ({D}^{\top})({D}^{\top})^{\dagger}\} \Sigma_{u|t}^{-1/2} \mu_{u|\Delta t} \|_2^2 \\
    &= \| w_a \{I - ({D}^{\top})({D}^{\top})^{\dagger}\}^2 \Sigma_{u|t}^{-1/2} \mu_{u|\Delta t} \|_2^2 \\
    &\leq \| w_a \{I - ({D}^{\top})({D}^{\top})^{\dagger}\} \|_2^2 \| \{I - ({D}^{\top})({D}^{\top})^{\dagger}\} \Sigma_{u|t}^{-1/2} \mu_{u|\Delta t} \|_2^2
\end{align*}
and, from \eqref{aGamma0},
\begin{align*}
    \| w_a \{I - ({D}^{\top})({D}^{\top})^{\dagger}\} \|_2^2
    &= \| a^{\top} \Gamma \|_2^2 - \| k_{ab} ({D}^{\top})^{\dagger} \|_2^2 \\
    &= k_{aa} - \| k_{ab} ({D}^{\top})^{\dagger} \|_2^2,
\end{align*}
where $k_{aa} := \| a^{\top} \Gamma \|_2^2$. To simplify this further, we use that
\begin{equation*}
    \| ({D}^{\top})^{\dagger} \|_2^2 = \left\| \frac{D}{k_{bb}} \right\|_2^2 = \frac{1}{k_{bb}^2} \|D\|_2^2 = \frac{k_{bb}}{k_{bb}^2} = \frac{1}{k_{bb}}.
\end{equation*}
Combining all of this gives us our final bound on $(III)$, which is given by
\begin{align*}
    |(III)|
    &\leq \| w_a \{I - ({D}^{\top})({D}^{\top})^{\dagger}\} \|_2 \| \{I - ({D}^{\top})({D}^{\top})^{\dagger}\} \Sigma_{u|t}^{-1/2} \mu_{u|\Delta t} \|_2 \\
    &\leq \sqrt{k_{aa} - \frac{k_{ab}^2}{k_{bb}}} \| \{I - ({D}^{\top})({D}^{\top})^{\dagger}\} \Sigma_{u|t}^{-1/2} \mu_{u|\Delta t} \|_2 \\
    &\leq \sqrt{k_{aa} - \frac{k_{ab}^2}{k_{bb}}} \| \Sigma_{u|t}^{-1/2} \mu_{u|\Delta t} \|_2.
\end{align*}
The final step is done because the $\| \{I - ({D}^{\top})({D}^{\top})^{\dagger}\} \Sigma_{u|t}^{-1/2} \mu_{u|\Delta t} \|_2$ term is unidentifiable, which means that different rotations of either the $\Gamma$ or $B$ matrices can lead to different values of this quantity. Putting all of these thoughts together gives us the following interval for the confounding bias of $a^{\top} Y$ in the presence of the negative controls: 
\begin{align*}
    & \frac{k_{ab}}{k_{bb}} b_j^{\top}\check{\mathcal{G}}_j M_{j}^{\dagger} \Sigma_{u|t}^{-1/2} \mu_{u|\Delta t} \quad \pm \nonumber \\
    & \Bigg( \Big| \frac{k_{ab}}{k_{bb}} \Big| \sqrt{k_{bb} - \| b_j^{\top}\check{\mathcal{G}}_j M_{j}^{\dagger} \|_2^2} \| (I - M_{j} M_{j}^{\dagger}) \Sigma_{u|t}^{-1/2} \mu_{u|\Delta t} \|_2 \nonumber \\
    & + \sqrt{k_{aa} - \frac{k_{ab}^2}{k_{bb}}} \ \|\Sigma_{u|t}^{-1/2} \mu_{u|\Delta t} \|_2\Bigg).
\end{align*}

Note that we obtain point identification of the treatment effect if (i) we have $m$ negative control contrasts ($c_j = m$), and (ii) either our negative control outcome is the same as the outcome of interest ($a = b_j$) or $a^{\top}\Gamma$ is colinear with $b_j^{\top}\Gamma$.
\end{proof}

\begin{proof}[Proof of Theorem 4]

Here we prove the results of Section 5.3.2 and show that under the conditions of Theorem 4, we can write the bias as a known function of identifiable quantities. As a reminder, the two pieces of information we have in order to identify the bias are 
\begin{align}
    a^{\top} \Gamma \Gamma^{\top} b &= K_{ab} \\
    b_j^{\top}\Gamma M_j &= b_j^{\top}\check{\mathcal{G}}_j \text{ for } j=1, \dots, J.
\end{align}
We prove the results separately by scenario, beginning with scenario 1: \\
\\
\textit{Scenario 1:} $a^{\top} \Gamma \in \mathcal{C}(\Gamma^{\top} b^*)$ and $b_j^{\top} \Gamma \in \mathcal{C}(M_j)$ for all $j \in \mathcal{J}^*.$ 

First, we need to verify that these two conditions are themselves verifiable despite only knowing $\Gamma$ up to rotation. The first condition holds because $a^{\top} \Gamma \in \mathcal{C}(\Gamma^{\top} b^*)$ holds if and only if $a^{\top} \Gamma R \in \mathcal{C}(R^{\top} \Gamma^{\top} b^*)$ for orthogonal matrices $R$. The second condition holds because we know the inner product between $b_j^{\top} \Gamma$ and every column of $M_j$. This implies that we can check the inner product of $b_j^{\top} \Gamma$ and any linear combination of the columns in $M_j$ and therefore we can check if $b_j^{\top} \Gamma$ lies in the span of $M_j$. The first condition implies that we can write
$$a^{\top} \Gamma = \sum_{j} w_j b_j^{\top} \Gamma$$
for some constants $w_j$ for $j=1, \dots J^*$ that are identifiable since we know the inner product of $a^{\top} \Gamma$ with $b_j^{\top} \Gamma$ for all $j$. Letting $M_{jk}$ be the $k$th column of $M_j$, the second condition implies that we can write
$$b_j^{\top} \Gamma = \sum_{k} d_{jk} M_{jk}^{\top}$$
for some constants $d_{jk}$ for $k=1, \dots, c_j$ that are identifiable because we know the inner product between $b_j^{\top} \Gamma$ and each of $M_{jk}$. Putting this together, we can see that the bias is written as
\begin{align*}
    a^{\top} \Gamma \Sigma_{u|t}^{-1/2} \mu_{u|\Delta t} &= \bigg( \sum_{j} w_j b_j^{\top} \Gamma \bigg) \Sigma_{u|t}^{-1/2} \mu_{u|\Delta t} \\
    &= \sum_j \sum_k w_j d_{jk} M_{jk}^{\top} \Sigma_{u|t}^{-1/2} \mu_{u|\Delta t},
\end{align*}
which is a fully identifiable quantity. This shows that the bias is a known function of identifiable quantities in this scenario.

\textit{Scenario 2:} There exists a $j$ such that $a^{\top} \Gamma$ is colinear with $b_j^{\top} \Gamma$ and $\Sigma_{u|t}^{-1/2} \mu_{u|\Delta t} \in \mathcal{C}(M_j)$.

Clearly we can check if $a^{\top} \Gamma$ is colinear with $b_j^{\top} \Gamma$ as this is invariant to rotation. Additionally, the second condition is straightforward to check as both $\Sigma_{u|t}^{-1/2} \mu_{u|\Delta t}$ and $M_j$ are identifiable quantities and therefore we can check if $\Sigma_{u|t}^{-1/2} \mu_{u|\Delta t}$ lies in the span of the columns of $M_j$. Because of these two conditions, we can write
\begin{align*}
    a^{\top} \Gamma &= w b_j^{\top} \Gamma \\
    \Sigma_{u|t}^{-1/2} \mu_{u|\Delta t} &= \sum_{k} d_{jk} M_{jk}
\end{align*}
for identifiable constants $w$ and $d_{jk}$ for $k = 1, \dots c_j$. This implies that the bias is identified as
\begin{align*}
    a^{\top} \Gamma \Sigma_{u|t}^{-1/2} \mu_{u|\Delta t} = w \sum_k d_{jk} b_j^{\top} \Gamma M_{jk},
\end{align*}
which is a fully identifiable quantity because we observe $b_j^{\top} \Gamma M_{jk}$ through the negative control assumption. 

\textit{Scenario 3:} $\Sigma_{u|t}^{-1/2} \mu_{u|\Delta t} \in \mathcal{C}(\Gamma^{\top} b^*)$ and $b_j^{\top} \Gamma \in \mathcal{C}(M_j)$ for all $j \in \mathcal{J}^*.$

Normally, it would not be possible to verify if $\Sigma_{u|t}^{-1/2} \mu_{u|\Delta t} \in \mathcal{C}(\Gamma^{\top} b^*)$ because $\Gamma$ is only known up to rotation, which can affect its corresponding column space. However, the second condition is verifiable since we know the inner product between $b_j^{\top} \Gamma$ and each column of $M_j$ from the negative control assumption, which allows us to check the inner product of $b_j^{\top} \Gamma$ with any linear combination of the columns of $M_j$. Given this second condition, we can write 
$$b_j^{\top} \Gamma = \sum_{k} d_{jk} M_{jk}^{\top},$$
for constants $d_{jk}$ for $k = 1, \dots, c_j$ that are identifiable. Because we can write $b_j^{\top} \Gamma$ in terms of identifiable quantities for all $j \in \mathcal{J}^*$, we can check the first condition to see whether $\Sigma_{u|t}^{-1/2} \mu_{u|\Delta t} \in \mathcal{C}(\Gamma^{\top} b^*)$. If true, then we can write 
$$\Sigma_{u|t}^{-1/2} \mu_{u|\Delta t} = \sum_j w_j b_j^{\top} \Gamma$$
for identifiable constants $w_j$ for $j \in \mathcal{J}^*$. This leads to writing the bias as
\begin{align*}
    a^{\top} \Gamma \Sigma_{u|t}^{-1/2} \mu_{u|\Delta t} &= a^{\top} \Gamma \sum_j w_j b_j^{\top} \Gamma \\
    &= \sum_{j} w_j K_{ab_j},
\end{align*}
which is identifiable from our factor confounding assumptions. 

\end{proof}

\subsection{Extension to multiple negative controls} \label{pf-thm-NCmultiple}

The previous results on negative controls only hold when there exists a single negative control pair, i.e., $J=1$. We used this example as a starting point to provide intuition for the benefits that negative control pairs can provide, but having more negative control pairs can reduce the width of the partial identification region even further. We extend these results to settings with $J \geq 1$ in Theorem \ref{thm-NCmultiple}. 

\setcounter{theorem}{4}

\begin{theorem} \label{thm-NCmultiple}
Under the assumptions of Theorem 3, suppose that we have $J$ negative control outcomes where $1 \leq J \leq m$, and $c_j \leq m$ negative control contrasts for each outcome. We define $b = [b_1, b_2, \dots, b_J]$, to be a $q \times J$ matrix that has each individual $b_j$ as its columns. Let $K_{ab} = a^{\top} \Gamma \Gamma^{\top} b$ be a $J$-dimensional row vector, and $K_{bb} = b^{\top} \Gamma \Gamma^{\top} b$ be a $J \times J$ matrix, both of which are identified from our factor model assumptions. Then the confounding bias, $\text{Bias}_{a,t_1,t_2}$, is in the interval
\begin{align} \label{NCinterval}
    & K^{*} D^{*} \Sigma_{u|t}^{-1/2} \mu_{u|\Delta t} \quad \pm \nonumber \\
    & \Bigg( \sum_{j=1}^J |(K^{*})_j| \sqrt{(K_{bb})_{j,j} - \| (D^{*})_{j \cdot} \|^2_2} \| (I - M_j M_j^{\dagger}) \Sigma_{u|t}^{-1/2} \mu_{u|\Delta t} \|_2 \nonumber \\
    & + \sqrt{K_{aa} - \| K^{*} b^{\top} \Gamma \|^2_2} \|  \Sigma_{u|t}^{-1/2} \mu_{u|\Delta t} \|_2 \Bigg),
\end{align}
where $(K^{*})_j$ is the $j$th element of $K^{*} := K_{ab} K_{bb}^{-1}$, $(K_{bb})_{j,j}$ is the $(j,j)$th entry of $K_{bb}$, and $(D^{*})_{j \cdot}$ is the $j$th row of a $J \times m$ matrix $D^{*}$, each row of which contains $b_j^{\top}\check{\mathcal{G}}_j M_j^{\dagger}$. 
\end{theorem}
The intuition for the partial identification region in Theorem \ref{thm-NCmultiple} is analogous to the case with a single negative control condition. Ideal negative control pairs are those with large confounding biases and whose treatment and outcome are affected by similar linear combinations of the unmeasured confounders as the treatment and outcome of interest, i.e., they have similar confounding mechanisms. We now give a proof of this result. 

\begin{proof}[Proof of Theorem \ref{thm-NCmultiple}]
Suppose that there are $J$ outcomes with at least one negative control treatment. Letting $\mathcal{Q} \subset \{1,\cdots,q\}$ denote a set of the indices of such outcomes and $\text{card}(\cdot)$ denote the cardinality of a set, we have that $J=\text{card}(\mathcal{Q})$. The negative control contrast provides us information about the plausible values of $b_j^{\top}\Gamma$ for $j \in \mathcal{Q}$. In particular, we can see the following. 
By Theorem 2 of \citet{penrose1955generalized}, a necessary and sufficient condition for (16) to have a solution is
\begin{equation*}
    b_j^{\top}\check{\mathcal{G}}_j M_{j}^{\dagger} M_{j} = b_j^{\top}\check{\mathcal{G}}_j,
\end{equation*}
in which case the general solution exists as
\begin{equation} \label{bGamma}
    b_j^{\top}\Gamma = b_j^{\top}\check{\mathcal{G}}_j M_{j}^{\dagger} + w_{b,j} (I - M_{j} M_{j}^{\dagger})
\end{equation}
for some arbitrary $m$-dimensional row vector $w_{b,j}$, where $M_{j}^{\dagger}$ denotes a generalized inverse of $M_{j}$, $M_{j} M_{j}^{\dagger}$ is the projection matrix onto the column space of $M_{j}$, and $M_{j}^{\dagger} M_{j}$ is the projection matrix onto the row space of $M_{j}$. 
Notice that \eqref{bGamma} satisfies
\begin{equation} \label{bGamma_decomp}
    \| w_{b,j} (I - M_{j} M_{j}^{\dagger}) \|^2_2 = \| b_j^{\top}\Gamma \|^2_2 - \| b_j^{\top}\check{\mathcal{G}}_j M_{j}^{\dagger} \|^2_2.
\end{equation}

For notational convenience, we denote the decomposition of \eqref{bGamma} by $b_j^{\top}\Gamma := D_j = D_j^{(1)} + D_j^{(2)}$ where $D_j^{(1)} := b_j^{\top}\check{\mathcal{G}}_j M_{j}^{\dagger}$ and $D_j^{(2)} := w_{b,j} (I - M_{j} M_{j}^{\dagger})$. It will also be useful later to break this matrix into two components since $D^{(1)}$ contains the estimable components and $D^{(2)}$ refers to the unidentified portion.
We refer to the stacked version of $D_j$'s as $\boldsymbol{D} = [D_1^{\top}, \dots, D_J^{\top}]^{\top}$, which is a $J \times m$ matrix. We can also define $\boldsymbol{b} = [b_1, b_2, \dots, b_J]$, which is a $q \times J$ matrix that has each individual $b_j$ as its columns. In other words, in the matrix form, we have that
\begin{align*}
    \left(\begin{array}{c}
    b_1^{\top}\Gamma \\ 
    \vdots \\ 
    b_J^{\top}\Gamma \\ 
    \end{array}\right)
    =
    \left(\begin{array}{c}
    b_1^{\top}\check{\mathcal{G}}_1 M_{1}^{\dagger} + w_{b,1} (I - M_{1} M_{1}^{\dagger}) \\ 
    \vdots \\ 
    b_J^{\top}\check{\mathcal{G}}_J M_{J}^{\dagger} + w_{b,J} (I - M_{J} M_{J}^{\dagger}) \\ 
    \end{array}\right)
    =
    \left(\begin{array}{c}
    D_1^{(1)} + D_1^{(2)} \\ 
    \vdots \\ 
    D_J^{(1)} + D_J^{(2)} \\ 
    \end{array}\right)
    =
    \left(\begin{array}{c}
    D_1 \\ 
    \vdots \\ 
    D_J \\ 
    \end{array}\right),
\end{align*}
i.e., $\boldsymbol{b}^{\top}\Gamma = \boldsymbol{D}^{(1)} + \boldsymbol{D}^{(2)} = \boldsymbol{D}$.

Under our factor modeling assumptions, we are able to identify
\begin{equation} \label{Kab}
    a^{\top} \Gamma \Gamma^{\top} \boldsymbol{b} := K_{ab}
\end{equation}
where $K_{ab}$ is a $J$-dimensional row vector.
Using the fact that $\boldsymbol{b}^{\top} \Gamma = \boldsymbol{D}$, we know that $a^{\top} \Gamma \boldsymbol{D}^{\top} := K_{ab}$, and thus all solutions to $a^{\top} \Gamma$ can be written as
\begin{equation} \label{aGamma_decomp1}
    a^{\top} \Gamma = K_{ab} (\boldsymbol{D}^{\top})^{\dagger} + w_a \left\{I - (\boldsymbol{D}^{\top})(\boldsymbol{D}^{\top})^{\dagger}\right\}
\end{equation}
for some arbitrary $m$-dimensional row vector $w_a$. 
This follows
\begin{equation} \label{aGamma_decomp}
    \left\| w_a \left\{I - (\boldsymbol{D}^{\top})(\boldsymbol{D}^{\top})^{\dagger}\right\} \right\|^2_2 = \| a^{\top} \Gamma \|^2_2 - \| K_{ab} (\boldsymbol{D}^{\top})^{\dagger} \|^2_2.
\end{equation}
Given that the rank of $\boldsymbol{D}$ is $J$, we can re-write $\boldsymbol{D}^{\dagger}$ as
\begin{equation*}
    \boldsymbol{D}^{\dagger} = \boldsymbol{D}^{\top} (\boldsymbol{D} \boldsymbol{D}^{\top})^{-1}
\end{equation*}
and $(\boldsymbol{D}^{\top})^{\dagger} = (\boldsymbol{D}^{\dagger})^{\top} = (\boldsymbol{D} \boldsymbol{D}^{\top})^{-1} \boldsymbol{D}$,
which implies that
\begin{align*}
    a^{\top} \Gamma &= K_{ab} (\boldsymbol{D} \boldsymbol{D}^{\top})^{-1} \boldsymbol{D} + w_a \left\{I - (\boldsymbol{D}^{\top})(\boldsymbol{D}^{\top})^{\dagger}\right\} \\
    &= K_{ab} (\boldsymbol{D} \boldsymbol{D}^{\top})^{-1} \boldsymbol{D}^{(1)} + K_{ab} (\boldsymbol{D} \boldsymbol{D}^{\top})^{-1} \boldsymbol{D}^{(2)} + w_a \left\{I - (\boldsymbol{D}^{\top})(\boldsymbol{D}^{\top})^{\dagger}\right\}.
\end{align*}
The bias for the estimand of interest is given by
\begin{align*}
    &~~~~a^{\top} \Gamma \Sigma_{u|t}^{-1/2} \mu_{u|\Delta t} \\
    &= K_{ab} (\boldsymbol{D} \boldsymbol{D}^{\top})^{-1} \boldsymbol{D}^{(1)} \Sigma_{u|t}^{-1/2} \mu_{u|\Delta t} + K_{ab} (\boldsymbol{D} \boldsymbol{D}^{\top})^{-1} \boldsymbol{D}^{(2)} \Sigma_{u|t}^{-1/2} \mu_{u|\Delta t} + w_a \left\{I - (\boldsymbol{D}^{\top})(\boldsymbol{D}^{\top})^{\dagger}\right\}  \Sigma_{u|t}^{-1/2} \mu_{u|\Delta t} \\
    &= \underbrace{K_{ab} K_{bb}^{-1} \boldsymbol{D}^{(1)} \Sigma_{u|t}^{-1/2} \mu_{u|\Delta t}}_{(I) \text{ bias correction}} + \underbrace{K_{ab} (\boldsymbol{D} \boldsymbol{D}^{\top})^{-1} \boldsymbol{D}^{(2)} \Sigma_{u|t}^{-1/2} \mu_{u|\Delta t}}_{(II)} + \underbrace{w_a \left\{I - (\boldsymbol{D}^{\top})(\boldsymbol{D}^{\top})^{\dagger}\right\}  \Sigma_{u|t}^{-1/2} \mu_{u|\Delta t}}_{(III)},
\end{align*}
where $K_{bb} := \boldsymbol{D} \boldsymbol{D}^{\top}$, and thus the first term $(I)$ is identifiable as our bias correction.
The remaining two terms, $(II)$ and $(III)$, are not identifiable and therefore we will bound them to create regions for the confounding bias that are compatible with the negative control conditions. Before bounding these terms, it is helpful to write $(II)$ as
\begin{align*}
    (II) = \sum_{j=1}^J \alpha_j w_{b,j} (I - M_{j} M_{j}^{\dagger}) \Sigma_{u|t}^{-1/2} \mu_{u|\Delta t},
\end{align*}
where $\alpha_j$ is the $j$th element of the $J$-dimensional row vector $K_{ab} (\boldsymbol{D} \boldsymbol{D}^{\top})^{-1} = K_{ab} K_{bb}^{-1}$. We know that, for $j=1,\cdots,J$,
\begin{align*}
    &~~~\left\{ \alpha_j w_{b,j} (I - M_{j} M_{j}^{\dagger}) \Sigma_{u|t}^{-1/2} \mu_{u|\Delta t} \right\}^2 \\
    &= \left\{ \alpha_j w_{b,j} (I - M_{j} M_{j}^{\dagger})^2 \Sigma_{u|t}^{-1/2} \mu_{u|\Delta t} \right\}^2 \\
    &= \| \alpha_j w_{b,j} (I - M_{j} M_{j}^{\dagger})^2 \Sigma_{u|t}^{-1/2} \mu_{u|\Delta t} \|^2_2 \\
    &\leq \alpha_j^2 \| w_{b,j} (I - M_{j} M_{j}^{\dagger})  \|^2_2 \| (I - M_{j} M_{j}^{\dagger}) \Sigma_{u|t}^{-1/2} \mu_{u|\Delta t} \|^2_2.
\end{align*}
The first equality holds because $(I - M_{j} M_{j}^{\dagger})$ is an idempotent matrix. Combining this with \eqref{bGamma_decomp},
\begin{align*}
    |(II)|
    &\leq \sum_{j=1}^J |\alpha_j| \| w_{b,j} (I - M_{j} M_{j}^{\dagger}) \|_2 \| (I - M_{j} M_{j}^{\dagger}) \Sigma_{u|t}^{-1/2} \mu_{u|\Delta t} \|_2 \\
    &= \sum_{j=1}^J |\alpha_j| \sqrt{\| b_j^{\top}\Gamma \|^2_2 - \| b_j^{\top}\check{\mathcal{G}}_j M_{j}^{\dagger} \|^2_2} \| (I - M_{j} M_{j}^{\dagger}) \Sigma_{u|t}^{-1/2} \mu_{u|\Delta t} \|_2,
\end{align*}
where $\| b_j^{\top}\Gamma \|^2_2 = (K_{bb})_{j,j}$ is the $(j,j)$th entry of $K_{bb}$, which is identifiable.

Now we need to provide a bound for the final term $(III)$ in the expression for the bias of interest. In a similar way, this can be done with \eqref{aGamma_decomp} as,
\begin{align*}
    |(III)|
    &\leq \left\| w_a \left\{I - (\boldsymbol{D}^{\top})(\boldsymbol{D}^{\top})^{\dagger}\right\} \right\|_2 \left\| \left\{I - (\boldsymbol{D}^{\top})(\boldsymbol{D}^{\top})^{\dagger}\right\} \Sigma_{u|t}^{-1/2} \mu_{u|\Delta t} \right\|_2 \\
    &= \sqrt{K_{aa} - \| K_{ab} K_{bb}^{-1} \boldsymbol{b}^{\top} \Gamma \|^2_2} \left\| \left\{I - (\boldsymbol{D}^{\top})(\boldsymbol{D}^{\top})^{\dagger}\right\} \Sigma_{u|t}^{-1/2} \mu_{u|\Delta t} \right\|_2 \\
    &\leq \sqrt{K_{aa} - \| K_{ab} K_{bb}^{-1} \boldsymbol{b}^{\top} \Gamma \|^2_2} \|  \Sigma_{u|t}^{-1/2} \mu_{u|\Delta t} \|_2,
\end{align*}
where $K_{aa} := a^{\top} \Gamma \Gamma^{\top} a$ is identifiable and $K_{ab} (\boldsymbol{D}^{\top})^{\dagger} = K_{ab} K_{bb}^{-1} \boldsymbol{b}^{\top} \Gamma$.
Putting all of this together, we can see that the confounding bias is in the region given by
\begin{align*}
    & K_{ab} K_{bb}^{-1} \boldsymbol{D}^{(1)} \Sigma_{u|t}^{-1/2} \mu_{u|\Delta t} \quad \pm \\
    & \Bigg( \sum_{j=1}^J |\alpha_j| \sqrt{\| b_j^{\top}\Gamma \|^2_2 - \| b_j^{\top}\check{\mathcal{G}}_j M_{j}^{\dagger} \|^2_2} \| (I - M_{j} M_{j}^{\dagger}) \Sigma_{u|t}^{-1/2} \mu_{u|\Delta t} \|_2 \\
    & + \sqrt{K_{aa} - \| K_{ab} K_{bb}^{-1} \boldsymbol{b}^{\top} \Gamma \|^2_2} \|  \Sigma_{u|t}^{-1/2} \mu_{u|\Delta t} \|_2 \Bigg),
\end{align*}
which is equivalent to
\begin{align*}
    & K^{*} \boldsymbol{D}^{*} \Sigma_{u|t}^{-1/2} \mu_{u|\Delta t} \quad \pm \\
    & \Bigg( \sum_{j=1}^J |(K^{*})_j| \sqrt{(K_{bb})_{j,j} - \| (\boldsymbol{D}^{*})_{j \cdot} \|^2_2} \| (I - M_{j} M_{j}^{\dagger}) \Sigma_{u|t}^{-1/2} \mu_{u|\Delta t} \|_2 \\
    & + \sqrt{K_{aa} - \| K^{*} \boldsymbol{b}^{\top} \Gamma \|^2_2} \|  \Sigma_{u|t}^{-1/2} \mu_{u|\Delta t} \|_2 \Bigg),
\end{align*}
where $K^{*} := K_{ab} K_{bb}^{-1}$ and $\boldsymbol{D}^{*} := \boldsymbol{D}^{(1)}$.
\end{proof}

\section{Algorithm for numerical approach in Section 5.3.1} \label{app-sec-NC-numerical}

We here outline the algorithm used to numerically determine the negative control partial identification regions. Note that we aim to minimize 
\begin{align*}
    \mathcal{F}(\widetilde{R})
    = (a^{\top} \widetilde{\Gamma} \widetilde{R} \Sigma_{u|t}^{-1/2} \mu_{u|\Delta t} - \beta)^2 + \sum_{j=1}^J \| b_j^{\top} \widetilde{\Gamma} \widetilde{R} M_{j} - b_j^{\top}\check{\mathcal{G}}_j \|_2^2,
\end{align*}
with
\begin{align*}
    \nabla_{\widetilde{R}} \mathcal{F}(\widetilde{R})
    &= 2 \left\{\widetilde{\Gamma}^{\top} a a^{\top} \widetilde{\Gamma} \widetilde{R} (\Sigma_{u|t}^{-1/2} \mu_{u|\Delta t}) (\Sigma_{u|t}^{-1/2} \mu_{u|\Delta t})^{\top} - \beta \widetilde{\Gamma}^{\top} a (\Sigma_{u|t}^{-1/2} \mu_{u|\Delta t}) \right\} + \\
    &~~~2\sum_{j=1}^J (\widetilde{\Gamma}^{\top} b_j b_j^{\top} \widetilde{\Gamma} R M_{j} M_{j}^{\top} - \widetilde{\Gamma}^{\top} b_j b_j^{\top}\check{\mathcal{G}}_j M_{j}^{\top}),
\end{align*}
where $\widetilde{R} \in \mathcal{V}_{m,m}$, the Stiefel manifold of all $m \times m$ orthogonal matrices. In order to run the algorithm, we first need to (i) specify $t_1$, $t_2$, and $a$ for the estimand of interest, and $\boldsymbol{b}=(b_1,\cdots,b_J)$ for the negative control contrasts, (ii) obtain $\widehat{\Gamma}$, $\widehat{\Sigma}_{u|t}$, $\widehat{\mu}_{u|\Delta t}$, $\check{g}(\cdot)$, $M_{j}$, and $b_j^{\top}\check{\mathcal{G}}_j$ for $j=1,\cdots,J$ using the estimation method described in Appendix \ref{sec:AppendixEstimationFactorModel}, (iii) define $\Theta = \{\theta_{a,t_1,t_2}: -1 \leq \cos(\theta_{a,t_1,t_2}) \leq 1\}$, a set of all candidate values of $\theta_{a,t_1,t_2}$, and (iv) specify a threshold value, $\delta$, as a selection criterion for $\theta_{a,t_1,t_2}$. Note that under the notation introduced in Section 5.3.1, $\delta = \delta_1 = \delta_2/J$. Then the algorithm can be summarized in Algorithm \ref{algo-gdaNC}. The idea is to find $\theta_{a,t_1,t_2}$ values that align with the conditions mentioned in Section 5.3.1. Once Algorithm \ref{algo-gdaNC} is implemented, we obtain $\Theta_{NC}$ and then construct the negative control partial identification region for bias as follows:
\begin{equation*}
    \text{Bias}_{a,t_1,t_2} \in \left\{ \|a^{\top}\Gamma\|_2 \| \Sigma_{u|t}^{-1/2} \mu_{u|\Delta t} \|_2 \cos(\theta_{a,t_1,t_2}): \theta_{a,t_1,t_2} \in \Theta_{NC} \right\}.
\end{equation*}

\setcounter{algocf}{1}
\begin{algorithm}[H]
\SetAlgoLined
\textbf{Input:} $t_1$, $t_2$, $a$ (estimand of interest), $\widehat{\Gamma}$, $\widehat{\Sigma}_{u|t}$, $\widehat{\mu}_{u|\Delta t}$ (estimates), $\boldsymbol{b}$, $M_{j}$, $b_j^{\top}\check{\mathcal{G}}_j$, $j=1,\cdots,J$ (negative controls), $\Theta$ (a set of candidate values of $\theta_{a,t_1,t_2}$), $\delta$ (threshold value).  \\
\textbf{Output:} $\Theta_{NC}$, a set of all selected values of $\theta_{a,t_1,t_2}$.

Initialize $\widetilde{R}_0$, by generating a random orthogonal matrix from the uniform distribution on the Stiefel manifold using the \texttt{rustiefel} function of the \texttt{rstiefel} package. \\
Initialize $\Theta_{NC} \leftarrow \varnothing$ (an empty set). \\
\For{each $\theta_{a,t_1,t_2} \in \Theta$}{

    Compute $\beta \leftarrow \|a^{\top}\widehat{\Gamma}\|_2 \| \widehat{\Sigma}_{u|t}^{-1/2} \widehat{\mu}_{u|\Delta t} \|_2 \cos(\theta_{a,t_1,t_2})$. \\
    
    Update $\widetilde{R}^* \leftarrow \arg\min_{\widetilde{R} \in \mathcal{V}_{m,m}} \mathcal{F}(\widetilde{R})$ using the \texttt{optStiefel} function of the \texttt{rstiefel} package, with $\widetilde{R}_0$ as the initial value and all unknown parameters replaced with their estimates. \\
    \If{$(a^{\top} \widehat{\Gamma} \widetilde{R}^* \widehat{\Sigma}_{u|t}^{-1/2} \widehat{\mu}_{u|\Delta t} - \beta)^2 \leq \delta$ \textbf{and} $\frac{1}{J}\sum_{j=1}^J \| b_j^{\top} \widehat{\Gamma} \widetilde{R}^* M_{j} - b_j^{\top}\check{\mathcal{G}}_j \|_2^2 \leq \delta$}
    {Add $\theta_{a,t_1,t_2}$ to the set $\Theta_{NC}$.}
    Set $\widetilde{R}_0 \leftarrow \widetilde{R}^*$.
}
   
\caption{Algorithm for finding negative control partial identification region}
\label{algo-gdaNC}
\end{algorithm}

\section{Estimation and inference on bounds} \label{sec:AppendixEstimationFactorModel}
In this section, we describe our estimation procedure for the required statistical models, as well as an inferential procedure that accounts for sampling uncertainty when working with partial identification regions or bounds. 
For now, we assume $\Sigma_u=I_m$, $\Sigma_{t|u}=\sigma^2_{t|u}I_k$, and $\Sigma_{y|t,u}=\sigma^2_{y|t,u}I_q$, though it is relatively straightforward to adapt these strategies to non-equal variances. 

\subsection{Estimation of factor model parameters}

The parameters corresponding to the treatment model, $\mu_{u|t}$ and $\Sigma_{u|t}$, can be estimated using factor analysis on the observed treatment matrix. From the treatment model, we have that
\begin{equation*}
    \text{Cov}(T)=BB^{\top}+\Sigma_{t|u},
\end{equation*}
where we want to estimate $B$ and $\Sigma_{t|u}$.
With a pre-specified number of factors, factor analysis can be applied to the covariance matrix of a standardized version of $T$, or the correlation matrix of $T$, which gives $B^*$ and $\Sigma_{t|u}^*$, from
\begin{equation*}
    \text{Cor}(T)=B^*B^{*\top}+\Sigma_{t|u}^*.
\end{equation*}
Here, $B^*$ is a matrix consisting of the factor loadings and ${\Sigma}_{t|u}^*=\text{diag}(\lambda_{t,1}^*.\cdots,\lambda_{t,k}^*)$ is a diagonal matrix whose entries are the uniquenesses of each standardized treatment, which can be obtained from standard statistical software such as the \texttt{factanal} function in R. Letting $\sigma^2_{t,j}$ indicate the variance of the $j$th treatment, and setting $W=\text{diag}(\sigma^2_{t,1},\cdots,\sigma^2_{t,k})$, we have that
\begin{align*}
    \text{Cov}(T) &= W^{1/2}\text{Cor}(T)W^{1/2} \\
    &= W^{1/2}B^*B^{*\top}W^{1/2} + W^{1/2}\Sigma_{t|u}^*W^{1/2} \\
    &= (W^{1/2}B^*)(W^{1/2}B^*)^{\top} + W^{1/2}\Sigma_{t|u}^*W^{1/2}.
\end{align*}
Hence, we can obtain estimates of $B$ and $\Sigma_{t|u}$ as follows:
\begin{align*}
    \widehat{B} &= \widehat{W}^{1/2}\widehat{B}^*, \\
    \widehat\Sigma_{t|u} &= \widehat\sigma^2_{t|u}I_k = \frac{1}{k} \sum_{j=1}^k \widehat\lambda_{t,j}^* \widehat\sigma_{t,j}^2 I_k.
\end{align*}
Note that it is straightforward to extend the estimate of $\Sigma_{t|u}$ to allow for unequal variances by using $\widehat\Sigma_{t|u} = W^{1/2}\Sigma_{t|u}^*W^{1/2}$, which we do in the analysis of PM$_{2.5}$ components. This provides the parameter estimates of the latent confounder model for a given $t$:
\begin{align*}
    \widehat{\mu}_{u|t}
    &= \widehat{B}^{\top}(\widehat{B}\widehat{B}^{\top}+\widehat{\Sigma}_{t|u})^{-1}t, \\
    \widehat{\Sigma}_{u|t}
    &= I_m - \widehat{B}^{\top}(\widehat{B}\widehat{B}^{\top}+\widehat{\Sigma}_{t|u})^{-1}\widehat{B}.
\end{align*}

The outcome model follows in a very similar way, though it requires an additional step where we first remove the effect of the treatment on the outcome. Specifically, we first estimate $\check{g}(\cdot)$ using a regression of $Y$ on $T$, and compute $Y-\check{g}(T)$. We then perform the same factor analysis steps as for the treatment model, but on the correlation matrix of $Y-\check{g}(T)$, which will provide an estimate $\widehat{\Gamma}$ of the unknown factor loadings. Once we have obtained estimates $\widehat{B}$ and $\widehat{\Gamma}$, we immediately obtain estimates of any identified quantity such as the partial $R^2$ values between the unmeasured confounders and treatment or outcome, as well as the bias bounds seen in Theorem 1. Note also that incorporating observed covariates $X$ is straightforward for both models. For the outcome model, it is simply included in the $\check{g}(\cdot)$ function, while for the exposures, we first find the residuals $T - E(T | X)$ where $E(T | X)$ is fit using a regression approach, and then do factor analysis on the residuals. 

One key component of factor analysis is determining how many factors to retain, as it is generally unknown in real applications. This is a well studied problem in the factor analysis literature, and there are a number of existing methods for determining the number of factors. Possibilities include the very simple structure criterion with complexity 1 and 2 \citep{revelle1979very}, the minimum average partial criterion \citep{velicer1976determining}, BIC, adjusted BIC, including the number of factors with eigenvalues greater than those of random data (parallel analysis), or the number of eigenvalues greater than 1. We leave discussion to the merits of each of these approaches to the existing literature on factor analysis, though empirically we have found both BIC and factor analysis to work reasonably well in the simulations explored in our work. We recommend favoring too many factors instead of too few factors as this will generally lead to more conservative inferences. In the analysis of PM$_{2.5}$ components, we used the largest number of factors allowable with 6 exposures and 6 outcomes, which is $m_1=3$ and $m_2 = 3$, which is also supported by the BIC values for the exposure and outcome factor models.

\subsection{Statistical inference on bounds}

Throughout the manuscript we have focused on population level quantities while ignoring the presence of sampling variability stemming from the estimation of unknown parameters. Standard inferential procedures used for point estimation do not immediately apply in this setting where we instead are targeting bounds for treatment effects. In this section, we describe a procedure based on the standard bootstrap in order to obtain upper and lower intervals for estimands of interest that have similar operating characteristics as standard confidence intervals in that they contain the true parameter $(1 - \alpha)100$\% of the time across repeated samples of the data. Suppose that we estimate all parameters of interest across $H$ bootstrap samples, where $H$ is large. This provides us with bootstrap samples of all unknown parameters denoted by $\widehat{\check{g}}^{(h)}(\cdot)$, $\widehat{\Gamma}^{(h)}$, and $\widehat{B}^{(h)}$ for $h=1, \dots H$. From these values of the model parameters, we can use the aforementioned procedures to obtain bounds for our estimand of interest, which we denote by $(\widehat{I}_l^{(h)}, \widehat{I}_u^{(h)})$ for $h=1, \dots H$. Given these bounds for each bootstrap sample, we can obtain the final bound $(\widehat{I}_l, \widehat{I}_u)$ to be the smallest bound such that $(1 - \alpha)100$\% of the individual bootstrap bounds, $(\widehat{I}_l^{(h)}, \widehat{I}_u^{(h)})$ are fully contained within the interval. 

Note that an analogous procedure could be done if the model parameters were all estimated within the Bayesian paradigm. For each posterior sample of the model parameters, we could calculate bounds, and then our final posterior bounds would be the smallest bounds that contain the desired proportion of the individual posterior sample bounds. 

\section{Justification for sampling on the Stiefel manifold} \label{sec:Stiefel}

\subsection{Theoretical justification}

Suppose we have $H$ randomly drawn rotation matrices $R^{(h)}$, for $h=1, \dots, H$. Define the following:
$$\rho_h = \sup_{R} \min_{j} d(R, R^{(j)}),$$
where $d(\cdot, \cdot)$ is the geodesic distance between two rotation matrices. This quantity shows what the maximum discrepancy is among all possible rotation matrices, when comparing the distance between the rotation matrix and it's closest rotation from resampling. Using the results in \cite{reznikov2016covering}, and noting that rotation matrices are manifolds of dimension $M(M-1)/2$ we can state that
$$\rho_h \asymp \left( \frac{\log H}{H} \right)^{\frac{2}{M(M-1)}}.$$
This shows that the maximum discrepancy between the rotation matrices tends to zero as $H \rightarrow \infty$, and that rotation matrices with higher dimensions require more samples $H$ to ensure that the space has been adequately explored. While this is useful, we ultimately care about the implications for the partial identification regions, which are a function of the rotation matrices themselves. Let $b(R)$ denote the confounding bias as a function of the rotation matrix. Further, let $R^*$ be the rotation matrix that leads to the largest bias in the sense that 
$$b(R^*) = \sup_{R} b(R).$$
Another way to describe $b(R^*)$ is that it corresponds to the upper endpoint of the desired partial identification region. Our goal is to show that we come arbitrarily close to this value through sampling uniformly across rotation matrices. Now let $R^{(j^*)}$ be the sampled rotation matrix that is closest to $R^*$.  An important fact is that $b(R)$ is a smooth function of the rotation matrix, and therefore it satisfies a Lipschitz continuity assumption, which allows us to state that
$$\left| b(R^*) - b(R^{(j^*)}) \right| \leq L d(R^*, R^{(j^*)}),$$
for some constant $L$. Given that $b(R^*) > b(R^{(j)})$ by definition, and that $d(R^*, R^{(j^*)}) \leq \rho_h$, we can see that
$$b(R^*) - b(R^{(j^*)}) \leq L \rho_h.$$
Given the rate condition above showing that $\rho_h \rightarrow 0$, this shows the desired result that the difference between our approximate upper bound and the true upper bound approaches zero. Note that a similar argument would hold for the minimum of the partial identification region as well. Lastly, note that some of our sampling algorithms require checking if certain conditions hold, such as negative control constraints or effect size constraints. The effective number of rotation matrices $H$ is therefore given by the number of rotation matrices that satisfy the given constraints. By similar arguments, because this is a number that we have full control of and can ensure is sufficiently large, this shows the validity of the proposed algorithm for finding partial identification regions even when constraints are imposed. 

\subsection{Impact of number of samples on partial identification region}

\begin{figure}[htbp]
    \centering
    \includegraphics[width=0.8\linewidth]{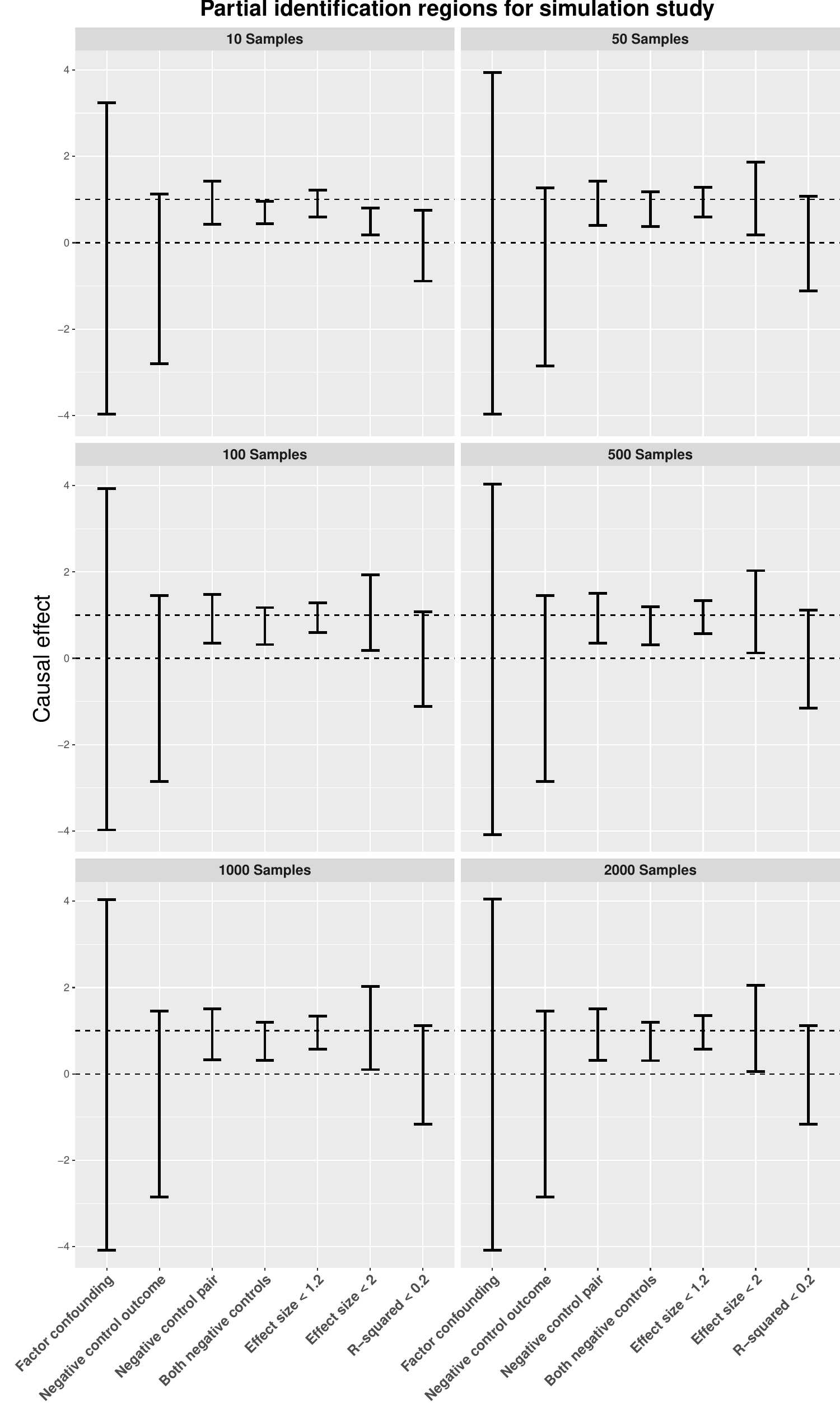}
    \caption{Partial identification regions for the simulation study as a function of the number of sampled rotation matrices. }
    \label{Fig:SimulationSamples}
\end{figure}

While the result in the previous subsection shows the theoretical validity of the proposed sampling algorithm, we now show empirically how it performs as a function of $H$, the number of sampled rotations. We highlight this in the context of the simulation study for which the true partial identification regions are known, but we have also seen that our partial identification regions for the Medicare data analysis stabilize rather quickly and do not require an exceedingly large number of sampled rotations. Figure \ref{Fig:SimulationSamples} shows the partial identification regions as a function of the number of rotations $H$ stored. We see that when there are only 10 sampled rotation matrices, the regions are not adequate as they have not stabilized, and at times do not contain the true values of the causal effect. The intervals are very close to stabilizing at 50 or 100 samples, though they are slightly smaller than the intervals with more samples for certain assumptions, such as the factor confounding only assumption. After 500 samples, however, there is no noticeable difference between the intervals. This suggests that a moderate number of samples is needed to obtain partial identification regions that are effectively identical to the regions one would obtain as $H \rightarrow \infty$. 

\section{Details of simulation study} 
\label{sec:AppendixSimDetails}

\begin{figure}[h]
    \centering
    \includegraphics[height=0.6\linewidth]{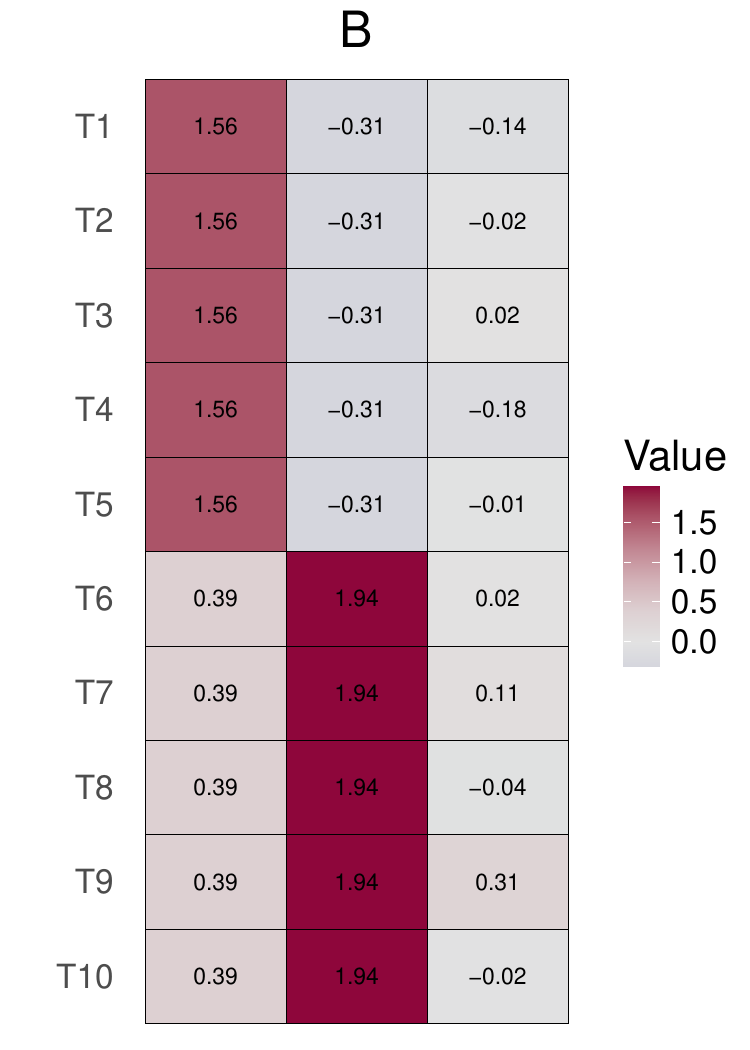}
    \includegraphics[height=0.6\linewidth]{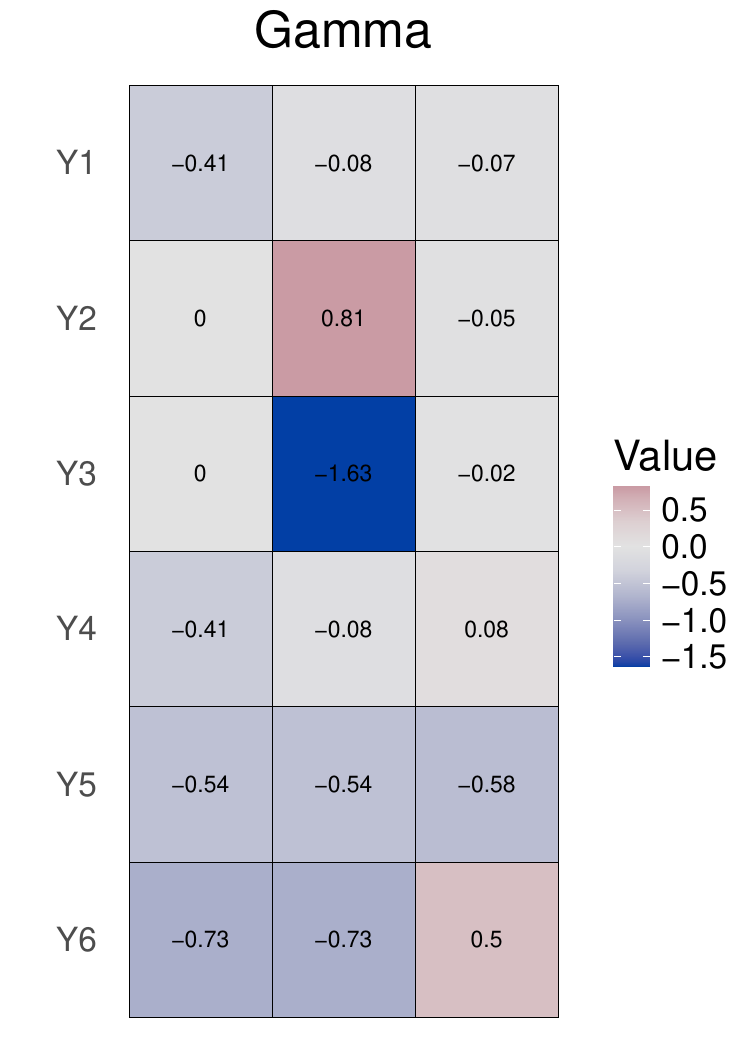}
    \caption{True values for the $B$ and $\Gamma$ matrices in the simulation study. }
    \label{Fig:SimulationFactors}
\end{figure}

\begin{figure}[h]
    \centering
    \includegraphics[height=0.7\linewidth]{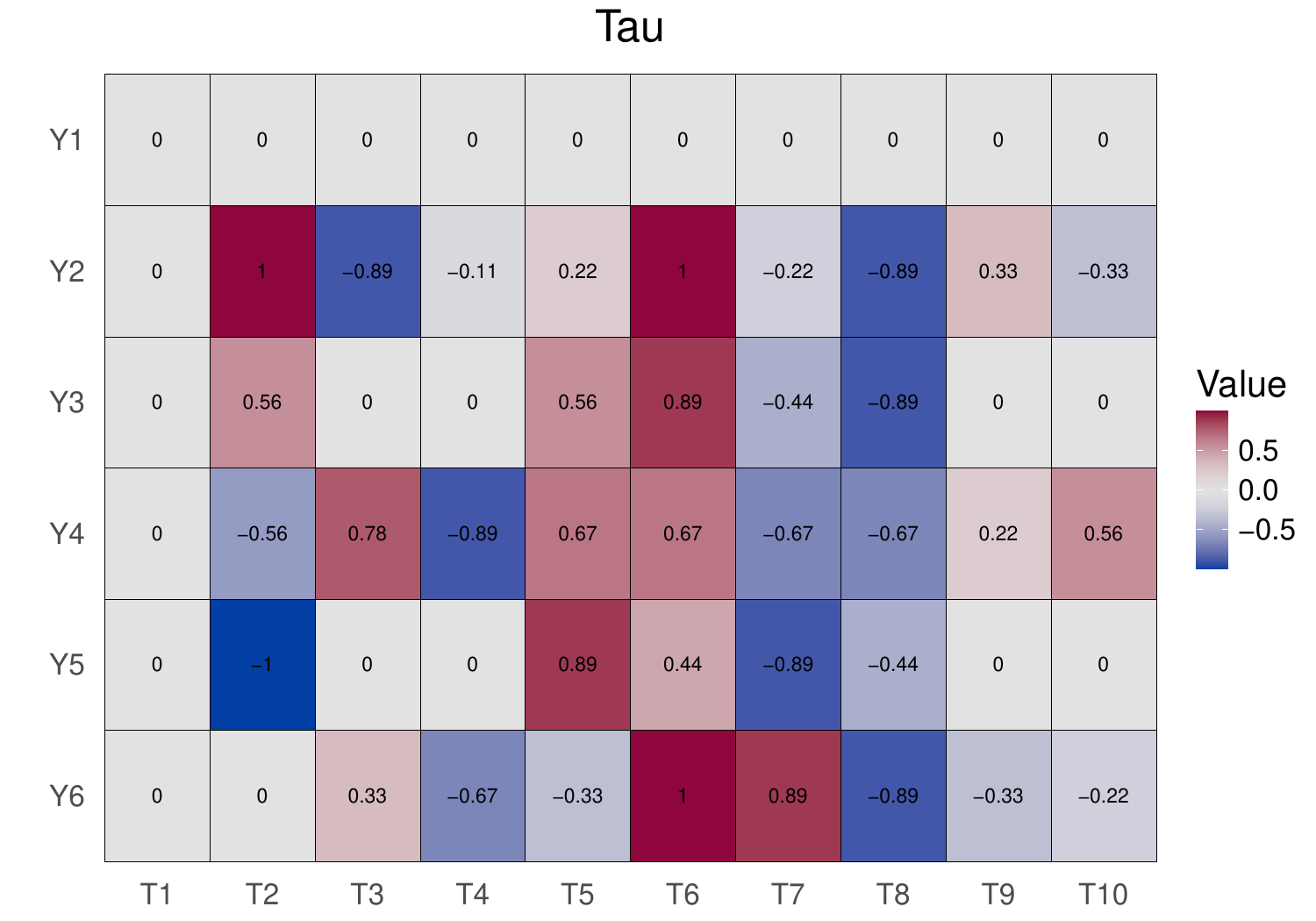}
    \caption{True values for the effect of each exposure on each outcome in the simulation study.}
    \label{Fig:SimulationTau}
\end{figure}

Here we provide more explicit details on the data generation mechanism used in the simulation study of Section 6. As mentioned in the manuscript, we use a single data set of size $n=10^6$ to effectively eliminate sampling variability and focus on uncertainty due to unmeasured confounding, and examine how partial identification regions change under varying assumptions. We set $m_1=3$ and $m_2 = 3$ though for the simulation we assume the same three variables for both treatment and outcome so that they are truly confounders and not predictors of the treatment or outcome only. For the dimension of the outcomes and treatments we set $q=6$ and $k=10$. The relevant covariance matrices have values $\Sigma_u = I_m$, $\Sigma_{t|u} = 2 I_k$, and $\Sigma_{y|t,u} = 270 I_q$. The true values of the coefficients relating the unmeasured variables $U$ to both treatments and outcomes can be found in Figure \ref{Fig:SimulationFactors}, where we standardize each row by dividing by the residual standard deviation for that treatment or outcome to ensure each row is on the same scale. We generate the outcome to be a linear function of the treatments, and the values of these coefficients can be found in Figure \ref{Fig:SimulationTau}. We also utilize a linear model for estimating the coefficients of the $\check{g}(\cdot)$ function, though the coefficients will be biased due to the presence of unmeasured confounding. The estimand of interest throughout is the effect of a one unit change for the 2nd treatment on the 2nd outcome. The first negative control outcome we use is the effect of the 2nd exposure on the first outcome, the negative control pair we use is the effect of the 3rd exposure on the 3rd outcome, and we use both of these negative controls for the assumption with two negative controls. 

A number of useful insights about the simulation design can be seen from these figures. For one, we can see why the negative control pair (3rd exposure on the 3rd outcome) is a better negative control than the first negative control outcome (2nd exposure on the 1st outcome) when aiming to estimate the effect of the 2nd exposure on the 2nd outcome. The 1st and 2nd rows of the $\Gamma$ matrix are nearly orthogonal, leading to a negative control that provides less information on the estimand of interest, while the 2nd and 3rd rows of $\Gamma$ have a large, negative inner product, leading to a larger reduction in the size of the partial identification region. The values of $B$ and $\Gamma$ also provide insight into the degree of unmeasured confounding bias. The partial R-squared values between each treatment and the unmeasured confounders range from 0.7 to 0.8, while the analogous partial R-squared values for the outcome range from 0.15 to 0.73. This shows that the simulation captures a situation with a large degree of potential confounding bias, which highlights the utility of the different assumptions (negative controls and effect size constraints) that are able to reduce the size of the partial identification region considerably despite the magnitude of the $B$ and $\Gamma$ coefficients. 

\section{Additional results in Medicare analysis}
\label{sec:AppendixExtraMedicare}

Here we present additional results on the effects of air pollutants on public health outcomes in the Medicare cohort. First we show the estimated factor loadings from both the exposure and outcome models, and the impact these values have on the partial R-squared between the unmeasured variables and each exposure and outcome. Then, we show additional results where we both 1) explore partial identification regions when we only assume factor confounding for the exposures, and 2) show results for exposures that were excluded from the main manuscript. 

\subsection{Factor loadings and implications for unmeasured confounding}

\begin{figure}[h]
    \centering
    \includegraphics[height=0.6\linewidth]{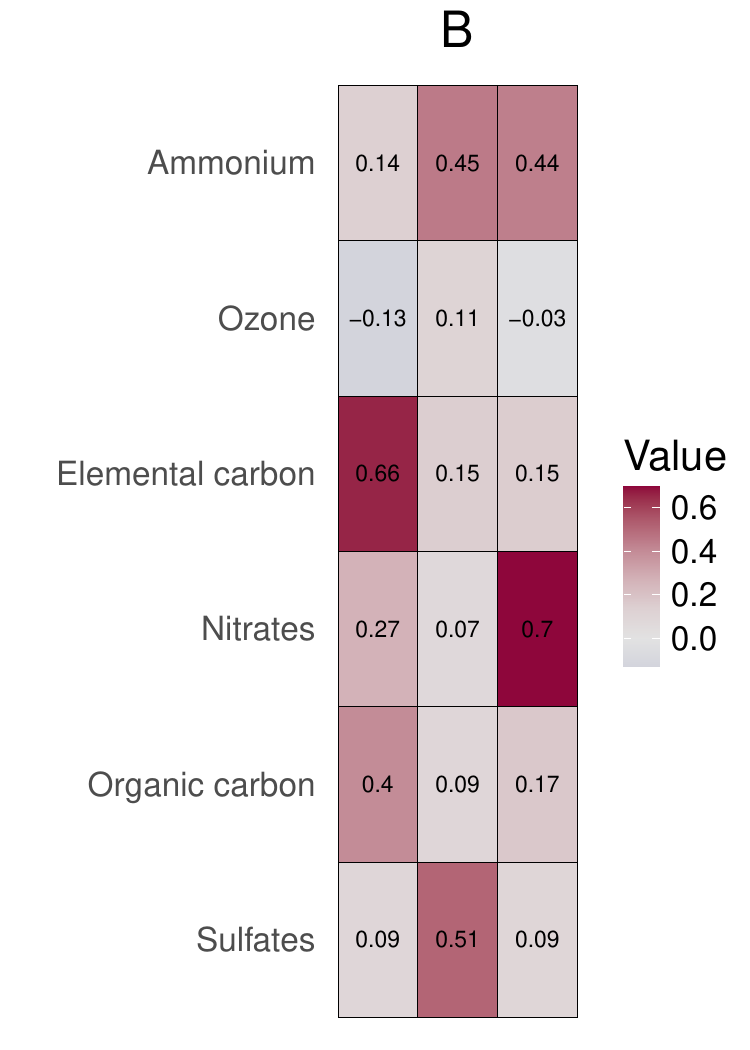}
    \includegraphics[height=0.6\linewidth]{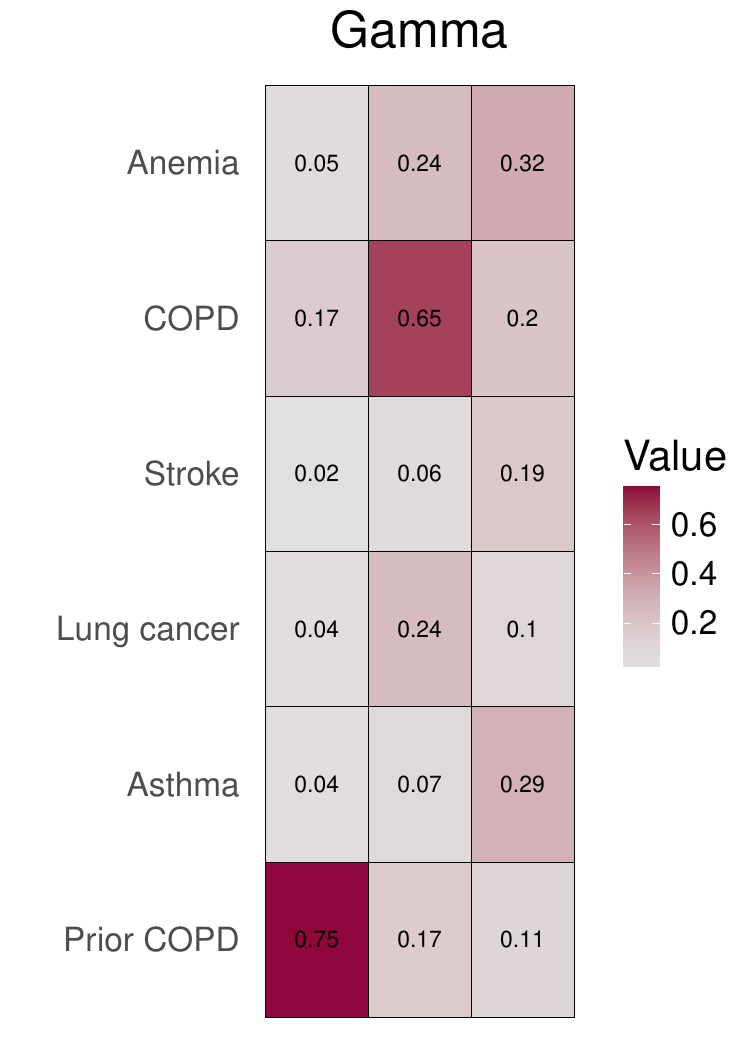}
    \caption{Estimates of the $B$ and $\Gamma$ matrices.}
    \label{Fig:MedicareFactors}
\end{figure}

Estimates of the factor model parameters are shown in Figure \ref{Fig:MedicareFactors}. The maximum number of factors that can be identified in factor analysis with 6 exposures and 6 outcomes (including the negative control outcome) is $m_1=3$ and $m_2 = 3$. We proceed with this choice for this reason, as it will provide the most conservative inferences about causal effects of interest. Additionally, we used BIC to select the number of factors and came to the same conclusion. We see that there are many large factor loadings, particularly in the exposure loading matrix, B.  However, we find that ozone has very small factor loadings, indicating little dependence with the other exposures. The first two columns of the outcome loading matrix, $\Gamma$,  largely reflect correlation among pulmonary outcomes: COPD, lung cancer and prior COPD. As such, prior COPD has the potential to be useful as a null control outcome for COPD and lung cancer. The factor loadings also provide information about $R^2_{a^{\top}Y \sim U|T, X}$ and $R^2_{d^{\top}T \sim U | X}$ for each outcome, $a$, and exposure, $d$ (see Table \ref{tab:MedicarePartialRsquared}). Larger values of these R-squared values imply wider partial identification regions for those exposures or outcomes due to the increased potential for unmeasured confounding bias.

\begin{table*}[ht]
\caption{Partial R-squared values between unmeasured variables and each of the exposures and outcomes}
\label{tab:MedicarePartialRsquared}
\begin{tabular}{@{}l llllll@{}}
  \hline
 & Anemia & COPD & Stroke & Lung cancer & Asthma & Prior COPD \\ 
 \hline
$R^2_{a^{\top}Y \sim U|T, X}$ & 0.236 & 0.777 & 0.044 & 0.075 & 0.102 & 0.990 \\ 
 \hline
 \hline
& Ammonium & Ozone & Elemental carbon & Nitrates & Organic carbon & Sulfates \\ 
\hline
 $R^2_{d^{\top}T \sim U | X}$ & 0.990 & 0.045 & 0.990 & 0.776 & 0.413 & 0.881 \\ 
   \hline
\end{tabular}
\end{table*}

\subsection{Only assuming factor confounding for the exposures}

In this section, we present results where we no longer assume factor confounding for the outcome, but continue to assume factor confounding for the exposures. Such an approach is closely related to work on single outcome and multi-treatment settings \citep{zheng2022bayesian}. Throughout, we still let $\Gamma$ denote the $q \times m$ matrix of coefficients that govern how the $m$ unmeasured variables affect the $q$ outcomes, but this parameter is now fully unidentified without the factor confounding assumption. Using the formulation in \cite{zheng2022bayesian} we can parameterize these coefficients for the outcome of interest in the following manner:
$$a^\top \Gamma = \sigma_{a'Y | T} \sqrt{R^2_{a'Y \sim U | T}} \Sigma_{u|t}^{-1/2} s,$$
where $s$ is a unit vector on the $(m-1)$-sphere. We have that $\sigma_{a'Y | T}$ is identifiable from the observed data, and $\Sigma_{u|t}^{-1/2}$ is identifiable under factor confounding for the exposures, but the other two terms are not identifiable. To move forward, we can first make an assumption about the potential magnitude of confounding bias, which is done by fixing $R^2_{a'Y \sim U | T}$. Then, we can explore partial identification regions for the causal effect of interest by randomly sampling $s$ uniformly over the $(m-1)$-sphere. An equivalent way to do this is to fix $s$ to be any unit-vector and then randomly rotate it by multiplying by a rotation matrix $R$ that we sample uniformly, just as we did in the implementation algorithm described in the manuscript.

\begin{figure}[h]
    \centering
    \includegraphics[width=0.8\linewidth]{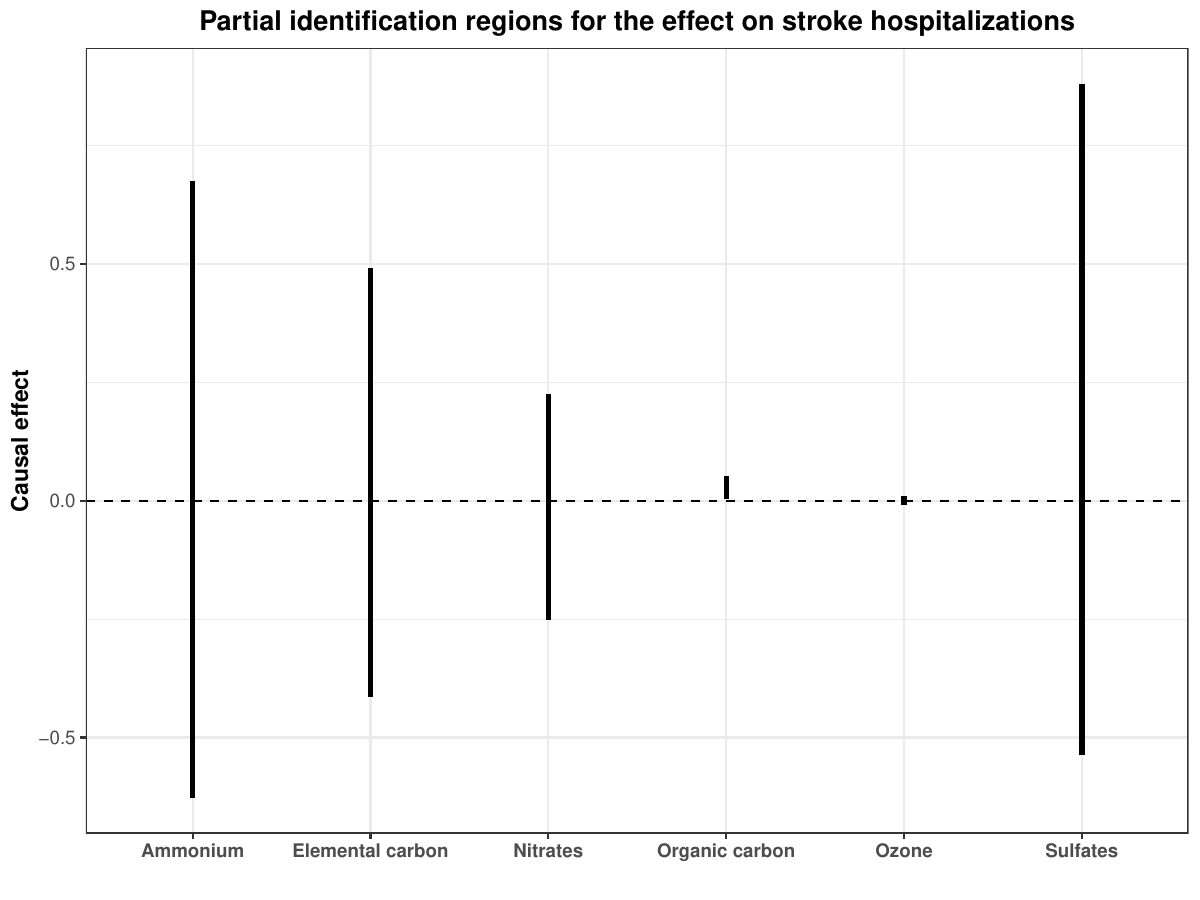}
    \caption{Partial identification regions under R-squared constraints for the effect of each exposure on stroke hospitalizations when only assuming factor confounding for the treatments. }
    \label{Fig:MedicareTreatmentOnly}
\end{figure}

\begin{figure}[!b]
    \centering
    \includegraphics[width=0.8\linewidth]{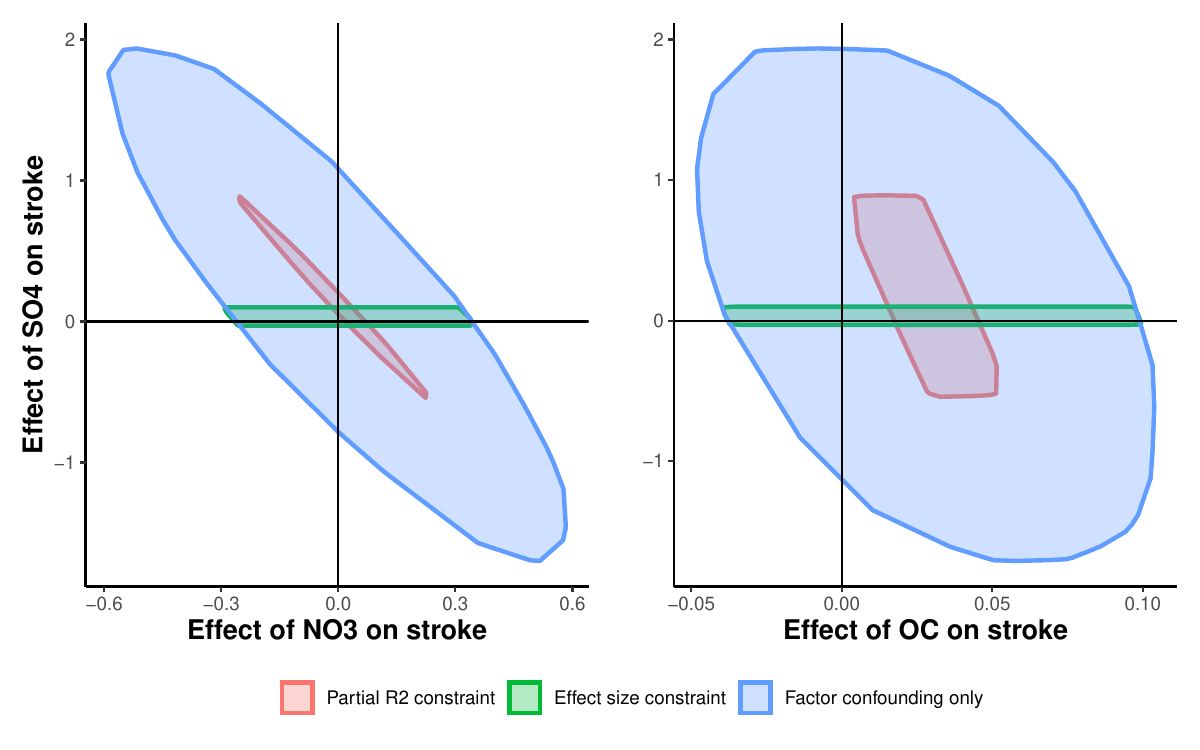}
    \caption{Bivariate partial identification regions under differing assumptions for the effect of organic carbon, nitrates, and sulfates on stroke hospitalizations when only assuming factor confounding for the treatments.}
    \label{Fig:MedicareTreatmentOnlyBivariate}
\end{figure}

We now apply this approach to the Medicare cohort, and we focus our results on the effects of all of the exposures on stroke hospitalization rates, which appeared to be affected by particulate matter in the analyses in the manuscript. We set $R^2_{a'Y \sim U | T} = 0.05$ as this corresponds to an association that is twice as strong as the strongest observed covariate found in our data. We then explore two additional assumptions that were used throughout the manuscript: 1) An effect size constraint ensuring that the effect of sulfates on the outcome is between $-$0.03 and 0.1, and 2) Assuming that $R^2_{d^{\top}T \sim U_{a,d}} < 0.05$ for all exposure contrasts. We present the univariate bounds under the second of these two assumptions in Figure \ref{Fig:MedicareTreatmentOnly}, and present bivariate bounds under either assumption in Figure \ref{Fig:MedicareTreatmentOnlyBivariate}. We exclude the effect size constraints from the univariate bounds as they are generally large and less informative, which makes it difficult to visualize the results under the R-squared assumption. From Figure \ref{Fig:MedicareTreatmentOnly} we can see that the bounds for all exposures contain zero, with the exception of organic carbon, which is fully contained above zero. This adds further support to the results of the manuscript showing that organic carbon affects stroke hospitalization rates. The bivariate plots in Figure \ref{Fig:MedicareTreatmentOnlyBivariate} show similar findings as in the manuscript, where there is a clear effect of organic carbon on stroke hospitalization rates, but also likely an effect of either sulfates or nitrates on stroke hospitalization rates as well. This can be seen in the left panel as the red partial identification region does not contain the null value of (0,0), which would correspond to no effect of either exposure. Lastly, both panels show that while the effect size constraint does drastically reduce the overall size of the partial identification region, it does not lead to informative bounds in this setting. 

\subsection{Effects on other exposures}

\begin{figure}[!b]
    \centering
    \includegraphics[width=0.8\linewidth]{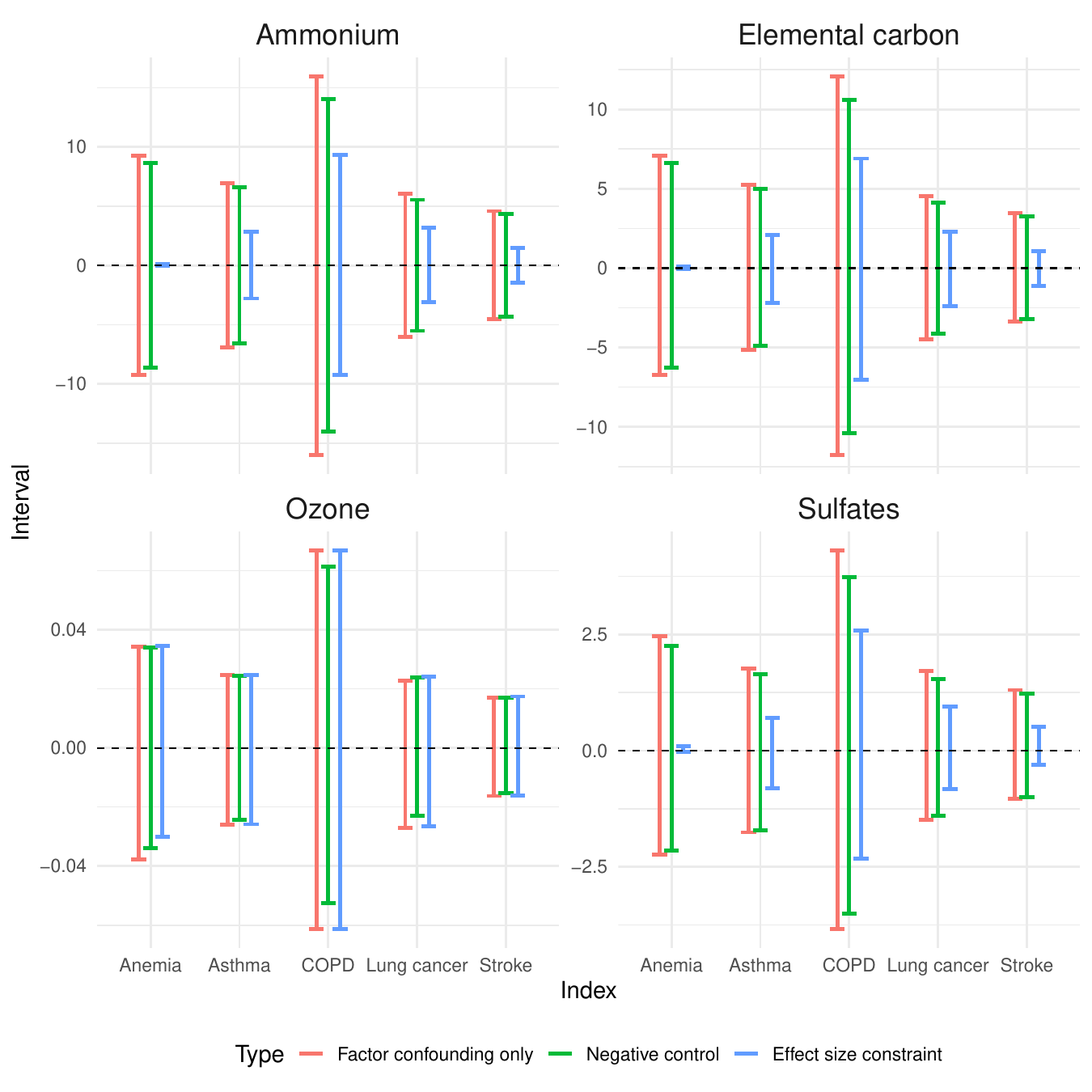}
    \caption{Partial identification regions for the causal effect of different exposures under the negative control and effect size constraints used in the manuscript.}
    \label{Fig:MedicareOtherExposures}
\end{figure}

In this section, we show results under the same constraints as in the manuscript in Section 8.2, but for the exposures that were not included in the manuscript. For this analysis the remaining exposures consist of elemental carbon, ammonium, ozone, and sulfates. The results can be found in Figure \ref{Fig:MedicareOtherExposures}. We see very wide partial identification regions for elemental carbon and ammonium, which is expected given the exceedingly large values of $R^2_{d^{\top}T \sim U | X}$ for these two exposures as seen in Table \ref{tab:MedicarePartialRsquared}. Similar results were found for sulfates, though the partial identification regions for this exposure are smaller in width. These findings show that it is unlikely within this framework to see informative partial identification regions under the shared confounding assumption when there is very strong dependence across exposures or outcomes. Nearly all of the unexplained variation in elemental carbon and ammonium are attributed to confounding in this situation, making it difficult to rule out large values of confounding bias. Ozone presents the opposite situation, as it has a very small value of $R^2_{d^{\top}T \sim U | X}$, and subsequently has very small partial identification regions. This points to the fact that the shared confounding assumption may be less plausible for ozone compared with the other exposures that have stronger dependence. 

\end{document}